\documentclass[a4paper,11pt]{amsart}
\usepackage{amssymb,amscd,amsxtra}
\usepackage{graphicx}
\usepackage{subcaption}

\usepackage{amsmath}
\usepackage{tabularx,booktabs,multirow}
\usepackage{xcolor}

\definecolor{red}{rgb}{1,0,0}

\usepackage[pdftex]{hyperref}
\hypersetup{%
bookmarksnumbered=true,%
colorlinks=true,%
setpagesize=false,%
pdftitle={},%
pdfauthor={}}
\usepackage{cleveref}

\newtheorem{Thm}{Theorem}[section]
\newtheorem{Lem}[Thm]{Lemma}
\newtheorem{Prop}[Thm]{Proposition}

\newtheorem{Def}[Thm]{Definition}
\newtheorem{Rem}[Thm]{Remark}
\newtheorem{Ex}[Thm]{Example}

\newcommand{\qi}{\mathbf{i}}
\newcommand{\qj}{\mathbf{j}}
\newcommand{\qk}{\mathbf{k}}
\newcommand{\eps}{\epsilon}

\newcommand{\N}{\operatorname{N}}
\newcommand{\arccot}{\operatorname{arccot}}
\newcommand{\Ann}{\operatorname{Ann}}

\makeatletter
\def\imod#1{\allowbreak\mkern10mu({\operator@font mod}\,\,#1)}
\makeatother

\title[Classification of higher mobility closed-loop linkages]{Classification of higher mobility closed-loop linkages}
\author{Tiago Duarte Guerreiro,  Zijia Li \and Josef Schicho}
\address{Department of Mathematical Sciences, Loughborough University, LE11 3TU, UK}
\email{t.guerreiro@lboro.ac.uk}
\address{KLMM, Academy of Mathematics and Systems Science, Chinese Academy of Sciences,
Beijing 100190, China}
\email{lizijia@amss.ac.cn -- Corresponding Author}
\address{Research Institute for Symbolic Computation, Johannes Kepler University, Altenberger Straße 69 A-4040 Linz, Austria}
\email{josef.schicho@risc.jku.at}
\keywords{Paradoxical Linkage, Mobility, Dual quaternion, Absolute Cone,  Bond Theory}
\date{}

\begin{document}

\begin{abstract}
		We provide a complete classification of paradoxical closed-loop $n$-linkages,  where $n\geq6$, of mobility $n-4$ or higher, containing revolute, prismatic or helical joints. We also explicitly write down strong necessary conditions for $nR$-linkages of mobility $n-5$. Our main new tool is a geometric relation between a linkage $L$ and another linkage $L'$ resulting from adding equations to the configuration space of $L$. We then lift known classification results for $L'$ to $L$ using this relation. 
		%	a variant $L'$ of it.  
		%linkages with different mobilities. 
		%which allows us to relate lower mobility linkages with higher mobility ones.
		%: From a linkage with high enough mobility we obtain lower mobility linkages by adding equations. A strong geometric relation between these is obtained and we use known classification results for our cases.
		%the freezing of a joint in a linkage with high enough mobility and its relation to the original unconstrained linkage.
		%a strong relation between a linkage and another one obtained by freezing a joint.
		% We also classify paradoxical 6-linkages of mobility 2 containing $P$ or $H$ joints.
\end{abstract}

\maketitle
\tableofcontents
%  \keywords{Dual quaternion \and Absolute Cone \and Bond theory \and Paradoxical linkage}
% \subclass{Primary 70B15; Secondary 14Q99\and 14H99\and 14J99\and 52C25}
\section{Introduction}

A closed-loop linkage is a mechanical device consisting of rigid bodies coupled together by joints and forming a single loop. Each joint allows for a different type of motion, namely, revolute (rotational around a fixed axis), prismatic (translational along a fixed direction) and helical (rotational and translational, where the rotation angle and the translation length are dependent). We call these $R$, $P$, and $H$ joints. A closed linkage with exactly $n$-joints is denoted by $n$-linkage. If all joints are of the same type, revolute, we call it an $n$R-linkage. 

Closed linkages are a crucial and basic object in the modern theory and applications of kinematics \cite{chen2015origami, Gosselin90, Mccarthy11book, Selig05book}.   They have been studied using algebraic and geometric methods in classical works such as Cayley \cite{Cayley1841}, Chebyshev \cite{Chebyshev1853theorie}, Sylvester \cite{Sylvester1874}, and Kempe \cite{Kempe1875general}.

The mobility (see \cref{def:mobility} for a precise description) of a closed-loop linkage is generically $n-6$ by the Chebychev-Gr\"ubler-Kutzbach criterion  \cite{kutzbach1929,Selig05book}, where $n$ is the number of $1$-degree-of-freedom (DOF) joints in the closed-loop. A closed-loop linkage whose mobility is higher than $n-6$  is called a paradoxical or overconstrained linkage. Mobile linkages with 4 joints of type $R$, $P$ or $H$ have been classified in \cite{Delassus1922chaines}. In \cite{Ahmadinezhad15}, the authors settle the classification problem for 5-linkages with at least one helical joint, which complements the classification of mobile linkages with five joints of type $R$, $P$ or $H$. Since the first mobile linkage with only revolute joints ($6R$-linkage) of  Sarrus~\cite{Sarrus53}, numerous paradoxical $6R$-linkages were discovered (see reviews in \cite{baker2002displacement, Baker03aj, Chen:r12, Dietmaier95,Li15phd}) but the classification of the paradoxical $6R$-linkages is still open. Furthermore, the classification of paradoxical linkages with $n\ge 6$ joints of type $R$, $P$ or $H$ is open too.

However, there has not been a systematic study yet for linkages with more than six joints or mobility higher than two. 
Our results constitute a significant step further in the solution to this problem. For a closed linkage,  if the motions of all joints are in a Sch{\"o}nflies subgroup of  $SE(3)$ which has dimension 4, then the linkage has mobility $n-4$, see \cite{Herve1978, Herve1999lie, Ravani03}. However, the converse is not true. For instance, the Goldberg 5R linkages of mobility one \cite{Goldberg} and 6R linkages of mobility two \cite{Hegedues13f}  are counter-examples. We completely classify higher mobility closed-loop $n$-linkages with mobility $n-4$.

%The classification of mobile linkages is an unsolved problem. 

%\cite{Chen:r12,Delassus1922chaines,Dietmaier95,Li2015}). 

%In particular, when such a linkage has exactly six revolute joints we expect no mobility at all. 
%There are, however, many examples of mobile linkages with $6$ (or even less) revolute joints (see \cite{Chen:r12,Delassus1922chaines,Dietmaier95,Li2015}) and these present many %interesting geometric properties 
% applications recently 
%in engineering (e.g.\ tensegrity bridges \cite{Pietroni2017}), 
%robotics (e.g.\ deployable mechanisms \cite{Zheng2016}), 
%material sciences (e.g.\ reconfigurable metamaterials \cite{Overvelde2017}), 
%medicine (e.g.\ auxetic stents \cite{Konakovic2018}), origami  \cite{wei2014origami,chen2015origami}, etc. 

% 

%In this paper, we take a step further in the solution of this problem.

\begin{Thm}[Main Theorem I, see Theorems \ref{classificationthm:6rm2}, \ref{p6result} and \ref{h6result}]
A 6-linkage of mobility 2 is one of the following cases:
\begin{enumerate}
	\item All joints are revolute with zero offsets and equal Bennett ratios; 
	\item There are exactly two $P$-joints and two pairs of neighbouring parallel $R$ or $H$-joints . \label{item:2}
	\item All axes of revolute or helical joints are parallel.
	%\item If there is at least one prismatic or helical joint not as in item \ref{item:2}, then all axes are parallel.
	\end{enumerate}
\end{Thm} 

%The full classification for $n-$linkages of mobility at least $n-4$, where $n>6$ and containing $$ is
\begin{Thm}[Main Theorem II, see Proposition \ref{prop:maxmob} and Theorems \ref{rnresult}, \ref{pnresult} and \ref{hnresult}]
The mobility of an $n$-linkage is at most $n-3$. %Moreover, an $n$-linkage, where $n>6$, has mobility at least $n-4$ only if all its revolute and helical axes are concurrent.
\begin{enumerate}
\item An $n$-linkage, where $n>6$, has mobility $n-4$ if and only if all its revolute and helical axes are parallel.
\item An $n$-linkage, where $n\geq 5$, has mobility $n-3$ only if all the axes of revolute or helical joints are concurrent.
\end{enumerate}
%All axes of an $nR-$linkage, $n>6$, of mobility at least $n-4$ are concurrent.
\end{Thm}

Moreover, we give necessary conditions that an $nR$-linkage of mobility $n-5$ satisfies:

\begin{Thm}[Main Theorem III, see Theorem \ref{thm:nrpartial}]
Every $n$R-linkage of mobility $n-5$ with $n>6$ has parallel neighbouring axes or triples of consecutive axes that satisfy a Bennett condition.
%On any $nR$-linkage, where $n>6$, of mobility $n-5$, there are parallel neighbouring axes or Bennett conditions.
\end{Thm}

Note that there are known $7R$-linkages of mobility two and $8R$-linkages of mobility three constructed using factorization of motion polynomials \cite{Liu21} and other synthesis methods \cite{Kong2015type, Pfurner2017algebraic}. A necessary condition when $n=6$ is also given in \cite{Li15q}. 

 The main new tool we use is the relation between a linkage $L$ with high enough mobility and a new linkage $L'$ obtained by adding equations to the configuration space of $L$. Each of these equations corresponds in practice to the \emph{freezing} of a joint in $L$ at a generic position, see  \cref{lem:fr}. Consequently, we reduce the mobility by one, which allows us to obtain classification results for $L$ by using previously known classification results for $L'$. Furthermore, the relation between $L$ and $L'$ is achieved by a new formulation of the Bennett and conditions for concurrency on revolute axes in terms of points in the absolute quadric cone, see \cref{lem:absconic}, which is interesting in itself. Finally, we use new considerations on bond theory for linkages with higher mobility, see \cref{brbt}, and an extension of the $abc-$lemma to dual quaternions, see \cref{lem:exabc}.

\section{Preliminaries}\label{ap}

 The real quaternions, $\mathbb{H}$, are the unique $4$-dimensional associative division algebra over $\mathbb{R}$. An element in $\mathbb{H}$ can be uniquely written as $p=p_0+p_1\mathbf{i}+p_2\mathbf{j}+p_3\mathbf{k}$, where $p_i \in \mathbb{R}$ and $\mathbf{i}^2=\mathbf{j}^2=\mathbf{k}^2=\mathbf{ijk}=-1$ are the usual Hamiltonian relations. Its conjugate is $\overline{p}:=p_0-p_1\mathbf{i}-p_2\mathbf{j}-p_3\mathbf{k}$. The algebra of quaternions is normed and the norm is a multiplicative function given by 
\begin{align*}
\N\colon \mathbb{H} &\longrightarrow \mathbb{R}\\
p &\longmapsto p\overline{p} = p_0^2+p_1^2+p_2^2+p_3^2.
\end{align*}

The quaternions are a good model for the 3D rotation group $SO(3)$, but it turns out that it is not enough to model the whole group of direct isometries of $\mathbb{R}^3$, $SE(3)$. For that, we need dual quaternions. Let $\mathbb{DH}$ be the $8$-dimensional associative algebra over $\mathbb{R}$ given by
\[
\mathbb{DH}:=\mathbb{H}[\epsilon]/(\epsilon^2).
\]

That is, an element in $\mathbb{DH}$ is of the form $h=p+\epsilon d$ where $p$ and $d$ are quaternions and $\epsilon^2=0$. The quaternions $p$ and $d$ are called the primal and dual parts of $h$, respectively. The dual quaternions are normed and the norm is given by 
  \begin{alignat*}{2}
  \N:\mathbb{DH}&\longrightarrow \mathbb{D} \\
  p+\epsilon d&\longmapsto (p+\epsilon d)\overline{(p+\epsilon d)}= p\overline{p}+\epsilon(p\overline{d}+d\overline{p}) \\
  &= p_0^2+p_1^2+p_2^2+p_3^2+2\epsilon(p_0d_0+p_1d_1+p_2d_2+p_3d_3),
\end{alignat*}
where $\mathbb{D}= \{a+b\epsilon \, | \, a,b \in \mathbb{R} \,\text{and}\, \epsilon^2=0 \}$. The subset of unit dual quaternions, $\mathbb{DH}^u$, given by elements with norm $1$, has a group structure.
%and is diffeomorphic to the tangent bundle of the unit sphere $S^3$, $TS^3 \cong S^3\times \mathbb{R}^3$.

The dual quaternions act on $\mathbb{R}^3$ and, in fact, restricting the action to the unit dual quaternions yields a surjection to $SE(3)$. Such a map is not injective, however, since any two unit dual quaternions differ by a sign map to the same isometry. Hence, it is natural to identify $h$ with any non-zero real multiple of $h$, therefore, get a one-to-one correspondence
\[
\mathcal{S}^*:= \{h \in \mathbb{P}^7 \, | \, \N(h) \in \mathbb{R}^* \} \leftrightsquigarrow SE(3).
\] 
%%%
Classically, it is usual to write $ \mathcal{S}^*$ as $\mathcal{S} \setminus E$ where 
\[
\mathcal{S} = \{(p_0:p_1:p_2:p_3:d_0:d_1:d_2:d_3) \in \mathbb{P}^7 \, | \, p_0d_0+p_1d_1+p_2d_2+p_3d_3=0 \}
\]
 is the Study Quadric. Inside $\mathcal{S}$ is the exceptional linear $3$-space
\[
E = \{(p_0:p_1:p_2:p_3:d_0:d_1:d_2:d_3) \in \mathbb{P}^7 \, | \, p_0=p_1=p_2=p_3=0 \} = \{\epsilon h\, | \, h \in \mathcal{S} \}.
 \]

The subset $\mathcal{S} \setminus E$ has a group structure inherited from the one in the unit dual quaternions. It is a classic result that $\mathcal{S} \setminus E$ is isomorphic to $SE(3)$. See \cite[Section~2.4]{husty10}.

In fact, the isomorphism can be given by $\psi : \mathcal{S} \setminus E \longrightarrow SE(3)$ such that 
\[
p + \epsilon d  \longmapsto \bigg(x \mapsto 1+\epsilon\frac{p\overline{d}-d\overline{p}+ px\overline{p}}{p\overline{p}}\bigg).
% p+\epsilon d \longmapsto (p+\epsilon d)(1+\epsilon x)(\overline{p}-\epsilon \overline{d}).
\]

Here, a point  $ x=(x_1,x_2,x_3)  \in \mathbb{R}^3$ is written in the form $1+\epsilon x = 1+x_1 \epsilon\mathbf{i}+x_2 \epsilon\mathbf{j}+x_3 \epsilon\mathbf{k}$.

\begin{Ex} The \cref{tab:simpletable} shows the correspondence between points in $\mathcal{S} \setminus E$ and isometries of $\mathbb{R}^3$.

\begin{table}[ht]
{\footnotesize
  \caption{Correspondence between elements in $\mathcal{S} \setminus E$ and $SE(3)$.}\label{tab:simpletable}
\begin{center}
 \begin{tabular}{m{5.6cm}m{5.6cm}} 
 \toprule
 Points in $\mathcal{S} \setminus E$ & Isometries in $\mathbb{R}^3$  \\ [0.5ex] 
 \midrule\midrule
 $(1:0:0:0:0:t_1/2:t_2/2:t_3/3)$ & Translation by $(t_1,t_2,t_3)$   \\ 
 \midrule
 $(t:u_1:u_2:u_3:0:0:0:0)$ where $t=\cot(\frac{\alpha}{2})$  & Rotation by $\alpha$ around the axis $u=(u_1,u_2,u_3)$ and $u_1^2+u_2^2+u_3^2=1$.   \\
 \bottomrule
\end{tabular}
\end{center}
}
\end{table}
\end{Ex}

Very often, we work in the complexification of the real algebra $\mathbb{DH}$. We can extend the dual quaternions to the algebra of complex dual quaternions, $\mathbb{DH}_{\mathbb{C}}$, defined as $\mathbb{DH} \otimes _{\mathbb{R}} \mathbb{C}$.

Inside $\mathbb{DH}_{\mathbb{C}}$ there is the subalgebra of complex quaternions, $\mathbb{H}_{\mathbb{C}}=\mathbb{H} \otimes _{\mathbb{R}} \mathbb{C}$. The following lemma says that it has a linear structure. 
\begin{Lem}
There is an algebra isomorphism,
\[
\mathbb{H}_{\mathbb{C}} \cong M_2({\mathbb{C}}).
\]
\end{Lem}
\begin{proof}
This is a special case of Wedderburn's theorem, see \cite{Wedderburn1908}, and follows directly from the fact that $\mathbb{H}_{\mathbb{C}}$ is a simple algebra over $\mathbb{C}$. \end{proof}

\begin{Lem}[$abc-$lemma] \label{lem:abc}
Suppose that $a$, $b$ and $c$ are norm zero complex quaternions. Then 
\[
abc=0 \,\, \implies ab=0 \quad \text{or} \quad bc=0.
\]
\end{Lem}

\begin{proof}
Let $X=\{x \in \mathbb{H}_{\mathbb{C}}\,| \, abx=0  \}$ and $Y=\{x \in \mathbb{H}_{\mathbb{C}} \,|\, bx=0  \}$. Clearly, $Y\subseteq X$. Both $X$ and $Y$ are right ideals of $\mathbb{H}_{\mathbb{C}}$ and, therefore, linear subspaces of $M_2(\mathbb{C})$. Suppose that $ab\not = 0$. Then $X \not =M_2(\mathbb{C})$ and $X$ and $Y$ are $2-$dimensional. Hence, $X=Y$. Since $c \in X$ by hypothesis, it follows that $c \in Y$. \end{proof}

%Let $p$ be a norm zero complex quaternion of the form $p=i-p_1$, where $p_1^2=-1$. Consider the right ideal $(i+p_1)\mathbb{H}_{\mathbb{C}}=\{x \in \mathbb{H}_{\mathbb{C}}\,| \, px=0\} \subseteq \mathbb{H}_{\mathbb{C}}$. 

%Suppose that $ab\not =0$. Let $X=\{x \in \mathbb{H}_{\mathbb{C}}\,| \, abx=0  \}$ and $Y=\{x \in \mathbb{H}_{\mathbb{C}} \,|\, bx=0  \}$. Then $Y\subseteq X$.

%By Wedderburn's theorem, both $X$ and $Y$ are $2$-dimensional linear subspaces of $M_2(\mathbb{C})$. Therefore, $X=Y$. Since $c \in X$ by assumption, it follows that $c \in Y$.  

The $abc-$lemma is one of the main tools in our results concerning linkages of mobility $2$ or higher. It was originally proven in \cite{Li15phd} to help with classifying the bond diagrams for $6R$-linkages with mobility $1$ with at most four connections in the sense of bond theory. We shall make extensive usage of this result. There is, in fact, a generalisation of the $abc-$lemma for complex dual quaternions that we will also use:

\begin{Lem}[Extended $abc-$lemma] \label{lem:exabc}
Suppose that $a$, $b$ and $c$ are norm zero complex dual quaternions, none of which is a quaternion multiple of $\epsilon$. Then $abc=0$ if and only if at least one of the following is true: 
\begin{enumerate}
	\item $ab=0$; 
	\item $bc=0$;
	\item $ab$ and $bc$ are both multiples of $\epsilon$. 
\end{enumerate}
\end{Lem}
\begin{proof}
It is clear that $ab=0$ or $bc=0$ implies $abc=0$. Suppose that both $ab$ and $bc$ are multiples of $\epsilon$. Then, for every $x \in \mathbb{DH}_{\mathbb{C}}$, we have that $abxbc=0$. We have thus to prove that there exists an element $x$ for which $bxb=b$. Any norm zero complex dual quaternion is of the form $u(i-\mathbf{i})v$ (by \cite[Lemma~1]{Hegedus2015theory}) where $u$ and $v$ are invertible dual quaternions and the $i$ is the complex imaginary unit. It follows that one only has to show that there is an element $x$ such that $(i-\mathbf{i})vxu(i-\mathbf{i})=(i-\mathbf{i})$. We take $x=\frac{1}{2i\N(u)\N(v)}\overline{(uv)}$.

To prove the converse we look at the left annihilator $\Ann_l$ of $bc$ in $\mathbb{DH}_{\mathbb{C}}$. If $bc$ is not a multiple of $\epsilon$ we have,
\begin{equation*}
a \in \Ann_l(bc)  = \mathbb{DH}_{\mathbb{C}} \overline{bc} 
% = \mathbb{DH}_{\mathbb{C}} \overline{c} \overline{b}
 \subseteq \mathbb{DH}_{\mathbb{C}} \overline{b} = \Ann_l(b),
\end{equation*}
which proves our claim.
\end{proof}

\section{Linkages}

Algebraically, we use the following definition of a linkage, which is taken from \cite{Ahmadinezhad15}:
\begin{Def}\label{def:linkage}
Let $n \in \mathbb{N}$. An \textbf{$n$-linkage} is a finite sequence of joints $(j_1,\ldots,j_n)$ where each joint is represented by an element in $\mathcal{S} \setminus E$ in the following way:
 \begin{description}
	\item[If $j_k$ is an $R$-joint] it is represented by the map $m_k : \mathbb{P}^1 \rightarrow \mathcal{S} \setminus E$ such that $t_k \longmapsto t_k-h_k$. The element $h_k$ is a fixed unit dual quaternion such that $h_k=p_k+\epsilon q_k$ and $h_k^2=-1$ and is of the form $h_k=(0:p_{k1}:p_{k2}:p_{k3}:0:q_{k1}:q_{k2}:q_{k3})$. Hence, $t_k-h_k=(t_k:-p_{k1}:-p_{k2}:-p_{k3}:0:-q_{k1}:-q_{k2}:-q_{k3})$. The parameter $t_k$ corresponds to the angle of rotation $2\arccot(t_k)$. 
	\item[If $j_k$ is a $P$-joint] it is represented by the map $m_k : \mathbb{P}^1\setminus \{0 \} \rightarrow \mathcal{S} \setminus E$ such that $t_k \longmapsto t_k-\epsilon p_k$. %The element $p_k$ is a fixed unit quaternion such that $p_k^2=-1$ and is of the form $p_k=(0:p_1:p_2:p_3:0:0:0:0)$. 
	Hence, $t_k-\epsilon p_k=(t_k:0:0:0:0:-p_{k1}:-p_{k2}:-p_{k3})$. The parameter $t_k$ corresponds to the extent of translation $\tfrac{\N(p_k)}{t_k}$. 
	%\item[If $j_k$ is a $C$-joint] it is represented by the map $m_k : \mathbb{P}^1 \setminus \{0 \} \times \mathbb{P}^1 \rightarrow \mathcal{S} \setminus E$ such that $(s_k,t_k) \longmapsto (s_k-\epsilon p_k)(t_k-h_k)$. The element $p_k$ is a fixed unit quaternion such that $p_k^2=-1$ and is of the form $p_k=(0:p_1:p_2:p_3:0:0:0:0)$ while $h_k=p_k+\epsilon d_k$ and $h_k^2=-1$.
	\item[If $j_k$ is an $H$-joint] it is represented by the map $m_k : \mathbb{R} \rightarrow \mathcal{S} \setminus E$ where $\alpha_k \longmapsto (1-\epsilon g_k \alpha_k p_k)(\cot(\frac{\alpha_k}{2})-h_k)$. The number~$g_k \in \mathbb{R}$ is a fixed constant called the pitch and %$h_k=p_k+\epsilon d_k$ is such that 
	$h_k^2=-1$. 
	\end{description}
	
If $L$ is a linkage, we impose that for each parameter $t_i$, the \textbf{loop-closure equation}, 
\begin{equation} \label{eq:loop}
m_1(t_1)\cdots m_n(t_n) \in \mathbb{R}\setminus \{ 0\}.
\end{equation}
is satisfied.

%the linkage needs those images (positions) of maps $m_1, \ldots, m_n$ that satisfying the \textbf{loop equation}, 
%\begin{equation} \label{eq:loop}
%m_1\cdots m_n \equiv 1.
%\end{equation}

The set of all  $(t_1,\ldots, t_n)$ satisfying the loop equation \eqref{eq:loop} for a linkage $L$ is called the \textbf{configuration set} of $L$, denoted by $K_L$. Hence, for instance, a linkage consisting of $6$ revolute joints has a configuration space in $(\mathbb{P}^1)^6$. Finally, the joins $j_k$ and $j_{k+1}$ are connected via a  mechanical part of the linkage which is called a \textbf{link} and is denoted by link $k+1$. For a closed-loop linkage, the joint $j_n$ and joint $j_1$ are connected via the link $1$.
%between each joint there is a \textbf{link}, that is, a mechanical part of the mechanism that connects the two joints.}

\end{Def}
\begin{Def}\label{def:mobility}
 The Zariski closure of all the complex configurations of an $n$-linkage is an algebraic variety which is denoted by $\overline{K_L}$, and its dimension is called the \textbf{mobility}.
\end{Def}
 With this dimension, we mean the complex dimension, and we are more interested in the special complex solutions in $\overline{K_L}$, which will be introduced in the next section. We say that a linkage $L$ is \textbf{mobile} if $\dim \overline{K_L} > 0$ and, in particular, we assume that no joint is frozen, that is, there is no $t_i$ for which the projection $\pi \colon \overline{K_L} \rightarrow \pi(\overline{K_L})$ to $t_i$ is zero-dimensional. Namely, we always consider the component of $\overline{K_L}$ with the highest dimension where no joint is frozen. Otherwise, we can treat the linkage as a reduced linkage with fewer joints.

\begin{Rem}
Although we consider the complex dimension for the configuration space of $L$, its real dimension is the same, provided that there is a real non-singular point in the highest complex dimensional components of the configuration space. If not, the real dimension is less than the complex dimension for the configuration space.
\end{Rem}

We explain how to associate several linkages to a given linkage $L$.
\begin{Def}[\cite{Ahmadinezhad15}]\label{lplrls}
Let $L$ be an $n$-linkage. 
\begin{itemize}
	\item We define the \textbf{spherical projection linkage} as the linkage $L_s$ obtained from $L$ by the map $\pi_s \colon \mathbb{DH} \rightarrow \mathbb{H}$ given by taking dual quaternions modulo $\epsilon$;
	\item If $L$ has at least one helical joint, we associate the linkage $L_r$ where every $H$-joint is replaced by an $R$-joint with the same axis. Similarly, we define $L_p$ where every $H$-joint is replaced by a $P$-joint with the direction parallel to the axis of the $H$-joint. 
\end{itemize}
\end{Def}
 
\begin{Rem} \label{rem:Lr}
By Theorem 9 in \cite{Ahmadinezhad15} and the paragraph right after that theorem, the mobility of $L_r$ (or $L_p$) is greater than or equal to the mobility of $L$.
\end{Rem}

\textbf{Denavit-Hartenberg Parameters and the Absolute Quadric Cone:}

We present geometric conditions of consecutive rotation joints. First, we introduce the set of Denavit-Hartenberg parameters of consecutive rotation joints. 
For indices $i=1,\dots,n$, let $l_i$ be the rotation axis of the $i$-th joint. 
The angle~$\alpha_i$ is defined as the angle of the direction vectors of
$l_i$ and $l_{i+1}$ (with some choice of orientation) and is called the \textbf{twist angle} between $l_i$ and $l_{i+1}$. We also set $c_i:=\cos(\alpha_i)$. %and $w_i=\cot(\frac{\alpha_i}{2})=\frac{cos(\alpha_i)+1}{sin(\alpha_i)}$.
 The number~$d_i$ is defined as the orthogonal distance of the lines $l_i$ and $l_{i+1}$ and is called the \textbf{twist distance} between $l_i$ and $l_{i+1}$.
Note that $d_i$ may be negative; this depends on some choice of the orientation of rotation axes. 
%which we denote by $n_i$. 
We set $b_i:=\frac{d_i}{\sin(\alpha_i)}$ as Bennett ratios (see \cite{mavroidis1995new}). Finally, we define the offset $o_i$ as the  signed distance of the two foot points on $l_i$. 
 %intersections of the common normals $n_{i-1}$ and $n_i$ with $l_i$. 

We recall two definitions \cite[Definition~3 and Definition~4]{Hegedues13b} as:
\begin{Def}[\cite{Hegedues13b}]
 For a sequence $h_i,h_{i + 1}, \ldots, h_j$ of consecutive joints, we define the coupling space $L_{i,i + 1, \ldots, j}$ as the linear subspace of $\mathbb{R}^8$ generated by all products $h_{k_1}h_{k_2}\cdots h_{k_s}$, $i \leq k_1 < \ldots < k_s \leq j$. (Here, we view dual quaternions as real vectors of dimension eight.) The empty product is included, its value is 1.
\end{Def}
\begin{Def}[\cite{Hegedues13b}]
 The dimension of the coupling space $L_{i,i + 1, \ldots, j}$ will be called the coupling dimension. We denote it by $l_{i,i + 1, \ldots, j} = \dim L_{i,i + 1, \ldots, j}$.
\end{Def}
Two rotation quaternions with the same axis are called \textbf{compatible}. A collection of rotation quaternions is called \textbf{concurrent} if their axes intersect at a common point or are parallel. Unless stated otherwise, we always assume our axes to be non-compatible. 

The following is a crucial geometric arrangement found by \cite{bennett03}.
\begin{Def}
Suppose $h_1,\,h_2$ and  $h_2,\,h_3$ are non-compatible rotation quaternions. We say that the triple $(h_1,h_2,h_3)$ satisfies \textbf{Bennett conditions} if the normal feet of $h_1$ and $h_3$ on $h_2$ coincide and $b_1=b_2$ or $b_1=-b_2$. In this case, we call $(h_1,h_2,h_3)$ a \textbf{Bennett triple}.
\end{Def}

%Even though the expected mobility of a $4R$-linkage is $-2$ by the Chebychev-Gr\"ubler-Kutzbach criterion, if three of the rotation quaternions satisfy Bennett conditions, there exists a uniquely defined fourth rotation quaternion forming a $4R$-linkage whose mobility is 1. 

Recall the following result:
\begin{Thm}[\cite{Hegedues13b}] \label{thm:bt1}
If $h_1, h_2, \ldots, h_n$ are rotation quaternions such that $h_i$ and $h_{i + 1}$ are not compatible for $i = 1, \ldots, n-1$, the following statements hold true:
\begin{itemize}
 \item All coupling dimensions $l_{1,\ldots,i}$ with $1 \leq i \leq n$ are even.
 \item The equation $l_{1,2} = 4$ always holds. Moreover, $L_{1,2} \subset \mathcal{S}$ if and only if the axes of $h_1$ and $h_2$ are concurrent.
 \item If $\dim L_{1,2,3} = 4$, then the axes of $h_1, h_2$ and $h_3$ are concurrent.
 \item If $\dim L_{1,2,3} = 6$, then the axes of $h_1, h_2$ and $h_3$ satisfy the Bennett conditions.%: the normal feet of $h_1$ and $h_3$ on $h_2$ coincide and the normal distances $d_{i}$ and twist angles $\alpha_{i}$ between consecutive axes are related by $b_1=\frac{d_{1}}{\sin \alpha_{1}} = \frac{\pm d_{2}}{\sin \alpha_{2}}=\pm b_2$.
\end{itemize}
\end{Thm}
\begin{Rem}
 Three axes intersecting at a point can be thought of as a degenerate Bennett triple, where the normal distances are zero. 
Three axes being all parallel  can be thought of as a degenerate triple, where the twist angles are zero.
\end{Rem}
%The next result derives analogous geometric conditions from four consecutive rotation joints $h_1$, $h_2$, $h_3$ and $h_4$.

%\begin{Lem} \label{lem:nobennett4}
%Let $h_1$, $h_2$, $h_3$ and $h_4$ be four consecutive rotation axes where no two neighbouring axes are parallel. It is impossible to have that there are infinitely many of rotation angles $\theta$ such that the new rotation axis $h_1^{\theta}$ obtaining with rotating $h_1$ around $h_2$ by $\theta$ fulfills Bennett conditions with $h_3$ and $h_4$, i.e., the normal feet of $h_1^{\theta}$ and $h_4$ on $h_3$ coincide and the normal distances and  twist angles between consecutive axes $h_1^{\theta}, h_3, h_4$ are related by $b_{\theta}=d_{1}^{\theta}/\sin \alpha_{1}^{\theta} = \pm d_{3}/\sin \alpha_{3}$. 
%\end{Lem}  
%\begin{proof}
% It is clear that there is only one parameter for determing the the normal distances and twist angles $d_{1}^{\theta},  \alpha_{1}^{\theta}$ which are rational fuctions of a univariate variable. Therefore they are constant when there are infinitely many of rotation angles such that the Bennett ratio is a constant. 
% For simplicity, we can lay the four axes on four parallel planes. We know that the $b_{\theta}$ is a constant when we rotate the $h_1$ around $h_2$ by 180 degrees. Then the first two planes containing $h_1$ and $h_2$ respectively coincide and they intersect at the foot point of $h_3$ on $h_2$. In addition, the $h_2$ and $h_3$ are parallel which contradicts our assumption.
%\end{proof}

We introduce a useful interpretation of the above results. Any plane in $\mathbb{R}^3$ is given by an equation $a_0x_0+a_1x_1+a_2x_2+a_3=0$ which is uniquely determined up to scalar multiplication. The coefficients of this equation are in one to one correspondence with points in ${\mathbb{P}^3}^{*}$, the dual projective space, except for the case when $a_0=a_1=a_2=0$ and $a_3=1$ which corresponds to the plane at infinity.

We define the dual absolute quadric cone $A \colon (a_0^2+a_1^2+a_2^2=0) \subset {\mathbb{P}^3}^{*}$. The set $A$ comes with an involution $\gamma \colon A \rightarrow A$ given by $p=(a_0:\ldots : a_3) \mapsto p^-=(\overline{a_0}: \ldots : \overline{a_3})$, where $\overline{a_i}$ denotes the complex conjugate of $a_i$ . Let $e=(0:0:0:1)$. Then $e \in A$ is the only real point of $A$, its only singular point and is the only point of $A$ fixed by $\gamma$.

Let $R=\{ h \in \mathbb{DH}\,\, | \,\, h^2=-1 \}$ be set of rotation quaternions. Then, $R$ also comes with an involution $\tau \colon R \rightarrow R$ given by $h \mapsto -h$, which coincides with quaternion conjugation. For every line $l$ in $\mathbb{R}^3$, there is a corresponding dual line $l^*$ in ${\mathbb{P}^3}^{*}$ which intersects $A$ in two non-real points. Notice that the revolute joints $h$ and $-h$ have the same fixed line of rotation. Then there is a 1 to 1 correspondence between the sets
\[
\{\text{Points in $A\setminus e$ modulo $\gamma$ }\} \longleftrightarrow \{\text{Lines in $\mathbb{R}^3$}\} \longleftrightarrow \{\text{Dual quaternions in $R$ modulo $\tau$}\}.
\]
We explain that this correspondence lifts to 
\[
\{\text{Points in $A\setminus e$}\} \longleftrightarrow \{\text{Oriented lines in $\mathbb{R}^3$}\} \longleftrightarrow \{\text{Dual quaternions in $R$}\}.
\]
The primal part of $h$ induces an orientation on the line of rotation. Hence a rotation quaternion corresponds to an oriented line in $\mathbb{R}^3$.  We now explain the first correspondence. Let $p=(a:b:c:d)$ and its complex conjugate $p^-$ be  points in $A\setminus e$ corresponding to a line $l$ in $\mathbb{R}^3$. To associate an orientation of $l$ for the point $p$, we take the point $(a,b,c) \in \mathbb{C}^3$. We write
\[
\begin{pmatrix}
a \\
b \\
c \\
d
\end{pmatrix}=
\begin{pmatrix}
a_1+a_2i \\
b_1+b_2i \\
c_1+c_2i \\
d_1+d_2i
\end{pmatrix}=
\begin{pmatrix}
a_1 \\
b_1 \\
c_1 \\
d_1
\end{pmatrix}+
\begin{pmatrix}
a_2 \\
b_2 \\
c_2 \\
d_2
\end{pmatrix}i.
\]
Let $u=a_1\mathbf{i}+b_1\mathbf{j}+c_1\mathbf{k}$ and $v=a_2\mathbf{i}+b_2+\mathbf{j}+c_2\mathbf{k}$. Since $p \in A \setminus e$, it follows that $a_1a_2+b_1b_2+c_1c_2 = 0$ and $u^2=v^2$. Therefore, $uv=(b_1c_2-c_1b_2)\mathbf{i}+(c_1a_2-a_1c_2)\mathbf{j}+ (a_1b_2-b_1a_2)\mathbf{k}$. The unit quaternion of $uv$ defines an orientation for the line $l$. Let $\lambda = \lambda_1+\lambda_2i \in \mathbb{C}^*$. Then $\lambda  (u+vi) = (\lambda_1u-\lambda_2v)+i(\lambda_2u+\lambda_1v)$ and 
\begin{align*}
(\lambda_1u-\lambda_2v) (\lambda_2u+\lambda_1v) &=  \lambda_1\lambda_2u^2-\lambda_1\lambda_2v^2 + \lambda_1^2uv-\lambda_2^2vu \\ 
&=\lambda_1^2uv -\lambda_2^2vu  \\
&=(\lambda_1^2 +\lambda_2^2)uv.
\end{align*}
Hence, the orientation is independent of the choice of representative in $\mathbb{P}^3$. Similarly, for $p^-$, we have the orientation defined by the unit vector of $-uv$. Moreover, we can associate an element of $R$ to the point $p$. Indeed, we associate to $p$ the element $h := u v - \epsilon (d_1v-d_2u)$. Notice that with a different representative of $p$, the dual quaternion $h$ only differs by a positive scalar. Hence, there is a positive scalar $\mu=\sqrt{u^2v^2}$ for which $\frac{h}{\mu} \in R$.     

%take the cross product $w := u \times v = (b_1c_2-c_1b_2,c_1a_2-a_1c_2, a_1b_2-b_1a_2)$, whose unit vector $(w_1,w_2,w_3)\in \mathbb{R}^3$ defines an orientation for the line $l$.  Let $\lambda = \lambda_1+\lambda_2i \in \mathbb{C}^*$. Then $\lambda \cdot (a:b:c) = (\lambda_1u-\lambda_2v)+i(\lambda_2u+\lambda_1v)$ and $(\lambda_1u-\lambda_2v) \times (\lambda_2u+\lambda_1v) = (\lambda_1^2+\lambda_2^2)u\times v$. Hence, the orientation  is independent of the choice of representative in $\mathbb{P}^3$. Similarly, for $p^-$, we have the orientation defined by the unit vector of $-w$. Moreover, we can associate an element of $R$ to the point $p$. Indeed, we associate to $p$ the element $h := u \times v + \epsilon (d_1v-d_2u)$.   

\begin{Lem} \label{lem:absrot}
The orbit of $h_1$ around $h_2$ is contained in the plane spanned by $p_{1},p_2,p_2^-$ in $A \setminus e$. 
\end{Lem}

\begin{proof}
To prove this claim, we may assume that $p_2=(1:i:0:0)$. 
Then the action of rotation by the angle $\alpha$ to $p_1=(x:y:z:w)$ gives 
  $(\cos(\alpha)x-\sin(\alpha)y:\sin(\alpha)x+\cos(\alpha)y:z:w)$ 
-- this is just the action of the transpose of the projective rotation matrix around the third coordinate axis. 
Hence the orbits are contained in planes with an equation $\mu z-\lambda w=0$ for some suitable $\lambda,\mu$, 
and these are planes through $p_2,p_2^{-}$. 
\end{proof}

\begin{Lem} \label{lem:absconic}
\begin{enumerate}
	\item The axes of joints $h_1$ and $h_2$ are concurrent if and only if the four points $p_1$,$p_1^-$,$p_2$ and $p_2^-$ are on a conic in $A$.
	\item The joints $h_1,\,h_2$ and $h_3$ form a Bennett triple with $b_1=b_2$ if and only if $p_1, \,p_2,\, p_2^{-},\,p_3$ are on a conic in $A \setminus e$. Similarly, $h_1,\,h_2$ and $h_3$  form a Bennett triple with $b_1=-b_2$ if and only if $p_1^{-}, \,p_2,\, p_2^{-},\,p_3$.
	\item If the joints $h_1,\,h_2,\,h_3$  form a Bennett triple and the joints $h_1,\,h_2,\,h_4$ form a Bennett triple,  then joints $h_4,\,h_2,\,h_3$ form a Bennett triple.
% 	\item The joints $h_1,\,h_2$ and $h_3$ satisfy the Bennett condition $b_1=b_2$ if and only if $p_1, \,p_2,\, p_2^{-},\,p_3$ are on a conic in $A \setminus e$. Similarly, $h_1,\,h_2$ and $h_3$ satisfy the Bennett condition $b_1=-b_2$ if and only if $p_1^{-}, \,p_2,\, p_2^{-},\,p_3$.
% 	\item If the joints $h_1,\,h_2,\,h_3$ satisfy a Bennett condition and the joints $h_1,\,h_2,\,h_4$ satisfy a Bennett condition,  then joints $h_4,\,h_2,\,h_3$ satisfy a Bennett condition.
\end{enumerate}
\end{Lem}

\begin{proof}

\begin{enumerate}
	\item 
The axes of $h_1$ and $h_2$ span a plane in $\mathbb{P}^3$. This plane is a point in the dual space contained in the dual lines $l_1^*$ and $l_2^*$. Hence, $l_1^*$ and $l_2^*$ also span a plane. The intersection of this plane with $A\setminus e$ is a conic that contains in particular $p_1$,$p_1^-$,$p_2$ and $p_2^-$. When the two axes are parallel the conic is degenerate and it consists of two lines passing through the vertex $e$.

\item We translate the lemma in coordinates assuming $b_1=b_2$. The case $b_1=-b_2$ is analogous. The statements remains unchanged if we apply an Euclidean isometry an all 3 lines $h_1,\,h_2$ and $h_3$ simultaneously. Therefore, we may assume w.l.o.g. that $h_2$ is the first axis and that the common normal between $h_1$ and $h_2$ is the third axis. The point $p_2$ corresponding to $h_2$ is $(0:1:i:0)$, and the
  point $p_2^{-}$ corresponding to $h_2$ with reversed orientation is $(0:1:-i:0)$. Assume $h_1$ corresponds to the point $p_1 = (x_1:y_1:z_1:w_1)$,
  and that $h_3$ corresponds to the point $p_3 = (x_3:y_3:z_3:w_3)$. Note that $x_1^2+y_1^2+z_1^2=x_3^2+y_3^2+z_3^2=0$. Because the common normal between $h_1$ and $h_2$ is the third
  axis, we can choose projective coordinates such that $x_1,\,y_1$ are
  purely imaginary (i.e. real multiplies of $i$) and $z_1$ is real. The orbit of rotation of $h_3$ around $h_2$ is contained in the plane
  spanned by $p_2,\,p_2^{-}$ and $p_3$ by \cref{lem:absrot}. Hence the first statement in the lemma is not changed
  if we apply a rotation around $h_2$ to $h_3$. Therefore, we may assume w.l.o.g. that the common normal between $h_2$ and $h_3$ is parallel to the third axis and that $x_3,\, y_3$ are purely imaginary and $z_3$ is real.

We now compute the angles between $l_2,\, l_1$ and $l_2,\,l_3$. W.l.o.g., we may assume $x_1^2+y_1^2=1,\, z_1=i$.
  If $\alpha_1$ is the angle between $l_2$ and $l_1$, then we have $x_1 = \sin(\alpha_1),\, y_1=\cos(\alpha_1)$.
  Similarly, if $\alpha_2$ is the angle between $l_2$ and $l_3$, then $x_3 = \sin(\alpha_2),\, y_3=\cos(\alpha_2)$.

  Computation of distances and the offset: the normal distance $d_1$
  between $h_2$ and $h_1$ is the imaginary part of $w_1$ divided by $i$.
  Similarly, the normal distance $d_2$ between $h_2$ and $h_3$ is 
  the imaginary part of $w_3$ divided by $i$.
  The real part of the $w_1$ is zero, and the real part of $w_3$ 
  is the offset $o$. Hence we have $w_1 = d_1i$ and $w_3 = d_2i+o$. Collecting everything about $p_1,\, p_2$ and $p_3$, we get:
	\begin{align*}
	 p_1 &= (\sin(\alpha_1):\cos(\alpha_1):i:d_1i)\\
   p_2 &= (0:1:i:0)\\
   p_3 &= (\sin(\alpha_2):\cos(\alpha_2):i:d_2i+o).
  \end{align*}
	Now its is clear that the first statement of the lemma is equivalent to
  \[
	x_1w_3-x_3w_1 = 
     \sin(\alpha_1)o+(\sin(\alpha_1)d_2-\sin(\alpha_2)d_1)i = 0
  \]
	and this is equivalent to the second statement.
\item This follows directly from the definition of Bennett triples. %Bennett conditions.
\end{enumerate}\end{proof}

\begin{Lem} \label{lem:fr}
Let $L$ be an $nR$-linkage with mobility $m\geq 2$ and joints $\{h_1, \ldots, h_n\}$. Suppose that freezing joint $h_n$ in a general position gives a linkage $L'$ such that $\{h_{n-1},h_1,h_2\}$ form a Bennett triple. %satisfy a Bennett condition. 
Then 
% \[
$\{h_{n-1},h_n,h_1,h_2 \}$
% \]  
are concurrent.
\end{Lem}

\begin{proof}
The rotation around $h_{n}$ is a one-parameter subgroup of Euclidean congruence transformations, which acts on the set of lines and therefore also on the absolute cone. Every orbit is contained in a plane passing through $p_n$ and $p_n^{-}$ by \cref{lem:absrot}. It follows that the orbit of $p_{n-1}$ is contained in the plane spanned by $p_{n-1},p_n,p_n^{-}$. For almost every point $x$ in the orbit,
we also have that the line corresponding to $x$ together with $h_1,h_2$ satisfies Bennett's condition. By \cref{lem:absconic}, this implies
that $x$ is contained in the plane spanned by $p_1,p_1^{-},p_2$. But the orbit of $p_{n-1}$ does span a unique plane, hence this unique plane
must contain the points $p_{n-1},p_n,p_n^{-},p_1,p_1^{-},p_2$.  Coplanarity of $p_n,p_n^{-},p_1,p_1^{-}$ is equivalent to the lines $h_n$ and $h_{n-1}$ being coplanar.  If three lines $(l_1,l_2,l_3)$ forms a Bennett triple, %fulfill a Bennett condition, 
 and $l_1,l_2$ are coplanar, then it follows all three lines are concurrent.
In our situation, this implies that all four lines $\{h_{n-1},h_n,h_1,h_2 \}$ must be concurrent.

\end{proof}

\section{Bond Theory} \label{brbt}
In this section, we define bonds associated to a linkage. We use it only for the results in \cref{sec:n-5}.
% and prove the Asymmetry Principle, \cref{lem:asy}. 
%Despite its simplicity, this is a far reaching concept that provides a framework for our analysis.  \textcolor[rgb]{1,0,0}{Correct}

Recall that a linkage is a sequence of joints $(j_1,\ldots,j_n)$, see \cref{def:linkage}.
\begin{Def}
Let $L$ be a closed-loop linkage and $S$ a subsequence of $L$.
\begin{itemize}
	\item We define the \textbf{length} of $S$, denoted by $|S|$, to be the number of elements of $S$;
	\item  We say that $S$ forms a \textbf{chain} if it is a finite sequence of consecutive joints of $L$.
\end{itemize}
 % Notice that $S$ is a chain if an only if $J\setminus S$ is a chain.
%Let $S\subseteq J$ be a non-empty set of joints of $L$. We say that $S$ forms a \textbf{chain} if it consists of a sequence of consecutive joints. Notice that $S$ is a chain if an only if $J\setminus S$ is a chain. %Moreover, we call $|S|$ the \textbf{length} of the chain $S$.
\end{Def}

%\begin{Rem}
%From a subset of joints, $S \subset J$, one can always form a chain associated to $S$, denoted by $S^c$, which is a smallest (with respect to number of elements) chain containing $S$. For each $S$ there are at most two possible subsets $S^c$.
%\end{Rem}

Let $L$ be the linkage $(j_1,\ldots,j_n)$ with $n$ joints. To a chain $S= (j_{r}, \ldots, j_{s} )$, where indices are taken $\mod n$, we associate the map $\Phi_S \colon K_L \longrightarrow \mathbb{DH}$ given by $m_{r}(t_r)\cdots m_{s}(t_s)$. We define the complementary chain to $S$, $\overline{S} = (j_{s+1},\ldots,j_{r-1})$ with indices taken modulo $n$. 
% By definition $\Phi_{\overline{S}} \colon K_L \longrightarrow \mathbb{DH}$ is the map satisfying 
The loop equation \cref{eq:loop} can be written as:
\[
\Phi_{S}(t)\Phi_{\overline{S}}(t) \in \mathbb{C}^* \,\, \forall \, t \in K_L.
\]
Moreover,
% \begin{align*}%
\[\overline{\Phi_{S}}(t)=
\overline{m_{r}(t)m_{r+1}(t)\cdots m_{s-1}(t)m_{s}(t)}.\]
% \end{align*}
% That is,
% \[
% \Phi_{\overline{S}}(t) = m_{s+1}(t)\cdots m_{r-1}(t).
% \]
% More explicitly,
% \begin{align*}
% \overline{\Phi_{\overline{S}}}(t)&=\overline{m_{r-1}}(t)\cdots \overline{m_{1}}(t)\overline{m_n}(t)\cdots \overline{m_{s+1}}(t) \\
% &=\overline{m_{s+1}(t)\cdots m_{r-1}(t)}.
% \end{align*}
% That is,
% \[
% \Phi_{\overline{S}}(t) = m_{s+1}(t)\cdots m_{r-1}(t).
% \]

The images of the maps $\Phi_S$ and $\Phi_{\overline{S}}$ correspond to the relative motion between the link $r$ and link $s+1$. We want to point out that it is also very interesting to look at what is the motion behavior at the compactification of $SE(3)$, see \cite{Hegedues13b}, i.e., the complex solutions of $\overline{K_L}$: We extend $\Phi_S$ to $\overline{K_L}$ and define the \textbf{set of bonds} of $L$, $B$, as 
\begin{equation*}
B = \{t \in \overline{K_L} \,\, |\,\, m_1(t) \cdots m_n(t)=0 \}.  
\end{equation*}

The following lemma, which is adapted from \cite{Hegedues13b}, is a consequence of the Affine Dimension Theorem \cite[Proposition~I.7.1,]{Hartshorne77book} and is the basic consequence of higher mobility from the perspective of bonds:

\begin{Lem} \label{cor:hart}
The set $B$ is an algebraic hypersurface of $\overline{K_L}$. Moreover, its complex dimension is one less than the mobility of the linkage.
\end{Lem}

 %\label{cor:hart}
Each bond component  of an $m$-dimensional irreducible component of $\overline{K_L}$ has dimension  $m-1$.  In this paper, we always consider such irreducible components where no joint is frozen.  There might be lower dimensional components in $\overline{K_L}$. 

\begin{Lem} \label{lem:curve}
Suppose $C \subset \mathbb{P}^1\times \mathbb{P}^1$ is a curve given by
\[
(i-h_s)(t_1-h_{s+1})(t_2-h_{s+2})(i-h_{s+3})=0
\]
where $h_i$ are non-parallel $R$-joints. Then, there is a Bennett triple between $(h_s, h_{s+1}, h_{s+2})$ or $(h_{s+1},h_{s+2},h_{s+3})$.
\end{Lem}

\begin{proof}
Since $C$ is a curve, there is a surjective map $C \rightarrow \mathbb{P}^1$ to one of the components, say, to the first one. In particular, the pre-image of $i \in \mathbb{P}^1$ is $(i,t)$ for some fixed $t \in \mathbb{P}^1$. That is, we have, over $i$,
\[
(i-h_s)(i-h_{s+1})(t-h_{s+2})(i-h_{s+3})=0.
\]
Let $a=i-h_s$, $b=i-h_{s+1}$, $c=(t-h_{s+2})(i-h_{s+3})$. Since $h_s$ and $h_{s+1}$ are not parallel, $ab\neq 0$ and $ab$ is not a quaternion multiple of $\eps$. By the extended abc-lemma \cref{lem:exabc}, it follows that $(i-h_{s+1})(t-h_{s+2})(i-h_{s+3})=0$. Hence,  we have $\dim L_{s+1,s+2,s+3} = 4$ or $\dim L_{s+1,s+2,s+3} = 6$. Since the axes are not parallel, by \cref{thm:bt1}, this is equivalent to a Bennett condition between $(h_{s+1},h_{s+2},h_{s+3})$.
% By \cref{lem:exabc}, it follows that $(i-h_{s+1})(t-h_{s+2})(i-h_{s+3})=0$, since $h_s$ and $h_{s+1}$ are not parallel. By \cref{thm:bt1}, this is equivalent to a Bennett condition between $(h_{s+1},h_{s+2},h_{s+3})$.
\end{proof}

%Define the subset of bonds $\widetilde{B}$ to be the ones of the form $(*,i,\beta_k,i,*)$ where $N(m_k(\beta_k))\not =0$.
%\begin{Lem} \label{lem:2sols}
%Let $\beta, \,\, \beta' \in \widetilde{B}$ be bonds such that $\beta_k \not = {\beta'}_k$. Let $S\subseteq J$ be the chain% of $R$-joints $\{j_{k-1},j_k,j_{k+1} \}$. If $\epsilon\Phi_S(\beta)=\epsilon\Phi_S(\beta')=0$, then $S$ is a set of %parallel joints.
%\end{Lem}                     
%\begin{proof}
%The equation $(i-p_{k-1})(t-p_k)(i-p_{k+1})=0$ is linear in $t \in \mathbb{P}^1$. Hence, if the coefficients are not all% identically zero, it has at most one solution. Since $\beta\not=\beta'$ are two solutions, it follows that the %coefficients of the equation are identically zero, that is,
%\[
%\Phi_S(\beta)=\Phi_S(\beta')=0 \quad \iff \quad \begin{cases}
%                    (i-p_{k-1})(i-p_{k+1})=0  \\
%                     (i-p_{k-1})p_k(i-p_{k+1})=0
%                 \end{cases}
%\]
%The first equation implies $h_{k-1} \parallel h_{k+1}$. From the second equation we then get $h_k \parallel h_{k+1}$.
%\end{proof}

The following definition is adapted from \cite{Hegedues13b,Li18survey}:
\begin{Def}
Let $\beta$ be a bond of $L$. Recall that $\beta$ is \textbf{attached} to a joint $j_k$ if $\N(m_k(\beta))=0$, where $m_k=t_k-\epsilon p_k$ if $j_k$ is a $P$-joint and $m_k=t_k-h_k$ if $j_k$ is an $R$-joint. We call $A_{\beta}$ the subsequence of joints to which $\beta$ is attached. Hence, $\beta \in B$ if and only if there is a $k$ such that $j_k \in A_{\beta}$.
\end{Def}

It was proven in \cite{Li18survey} that the mobility of a specific joint in a linkage is equivalent to the existence of a bond attached to that joint. Hence, for a mobile linkage, there is always some $\beta$ for which $A_{\beta}$ is non-empty. In fact, it is known that $|A_{\beta}| \geq 2$ in general. See \cite{Hegedues13b}. 
% In addition, in \cite[Proposition 1.]{Hegedues13b}, it is also proved that the dimension of the set of all bonds is $m-1$ if the linkage has  mobility $m$. 
Furthermore, we have the following result:

\begin{Lem} \label{lem:chain}
Let $L$ be an $nR$-linkage and suppose that $A_{\beta}$ is a chain. Then there is at least one pair of parallel joints.  
\end{Lem}

\begin{proof}
Without loss of generality we can assume $A_{\beta}=(j_1,\ldots,j_r)$. Then, 
\[
\Phi_{A_{\beta}}(\beta)=(i-h_1)\cdots (i-h_r)=0,
\]
where $h_i=p_i+\epsilon q_i$ is a rotation dual quaternion, that is, $h_i^2=-1$. Multiplying by $\epsilon$ we get $(i-p_1)\cdots (i-p_r)=0$. By repeated usage of the $abc$-lemma there is an index $s$ for which $(i-p_s)(i-p_{s+1})=0$ which implies that $p_s$ and $p_{s+1}$ are parallel.
\end{proof}

\begin{figure}[tbhp]
  \centering
  \includegraphics[width=3.04cm]{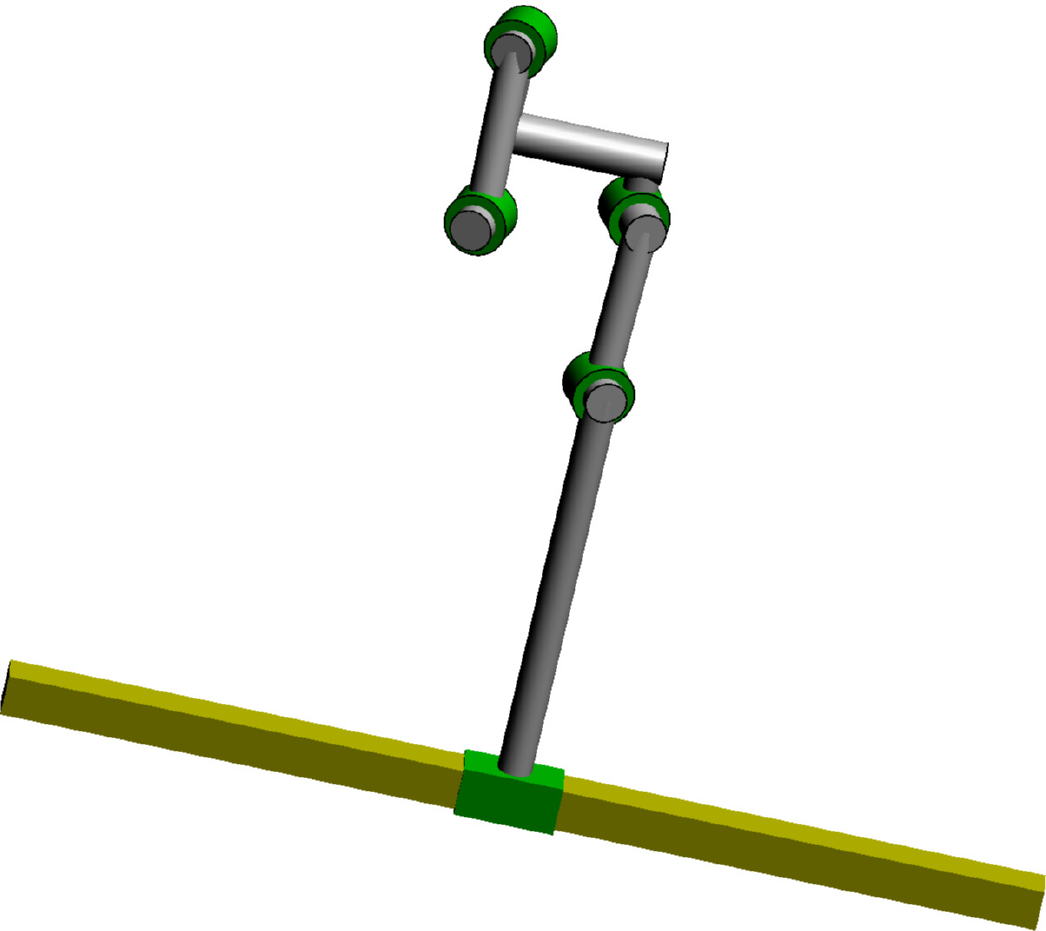}
  \includegraphics[width=3.03cm]{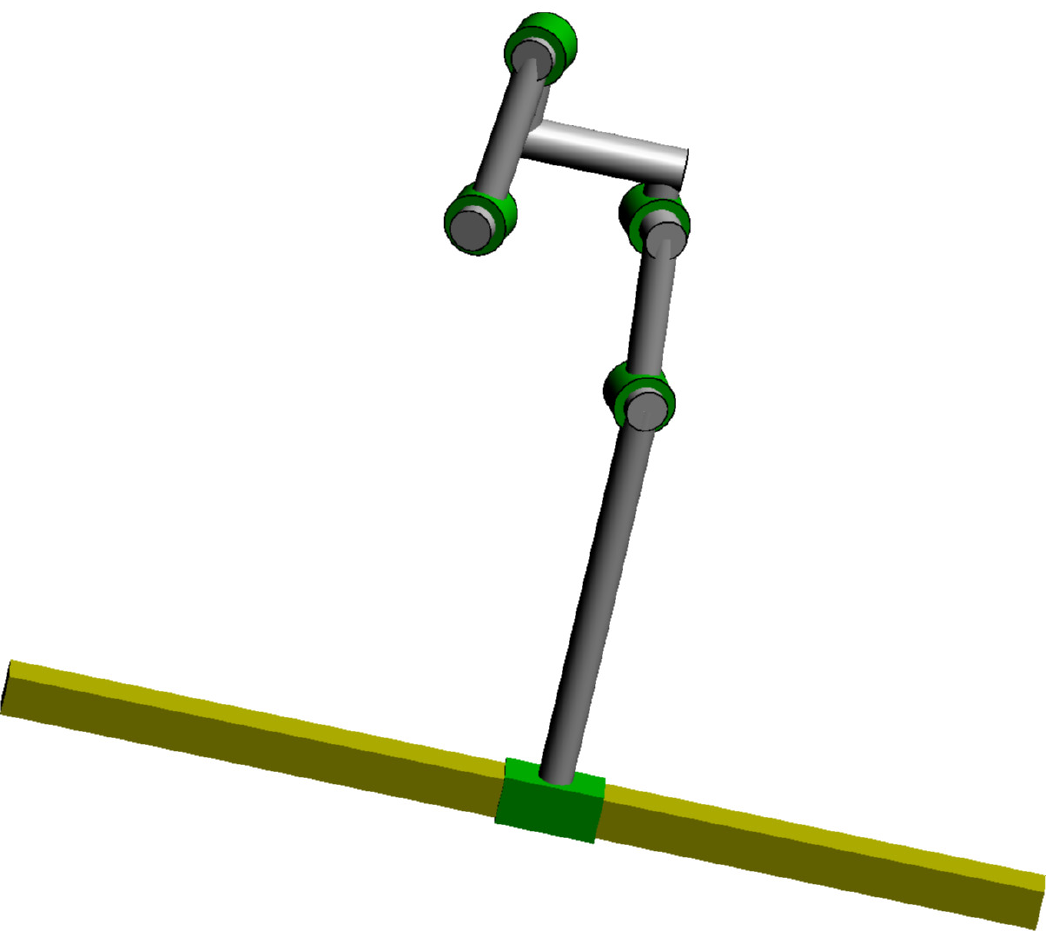}
  \includegraphics[width=3.03cm]{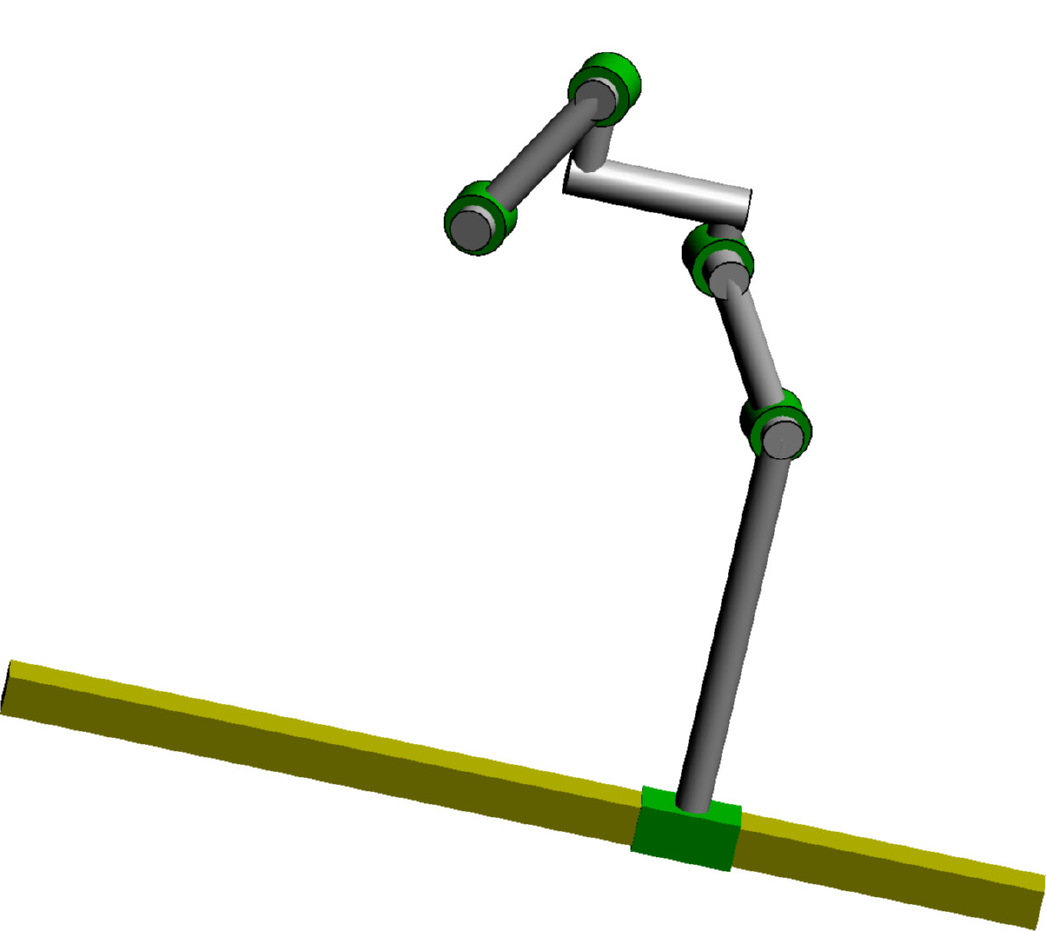}
  \includegraphics[width=3.03cm]{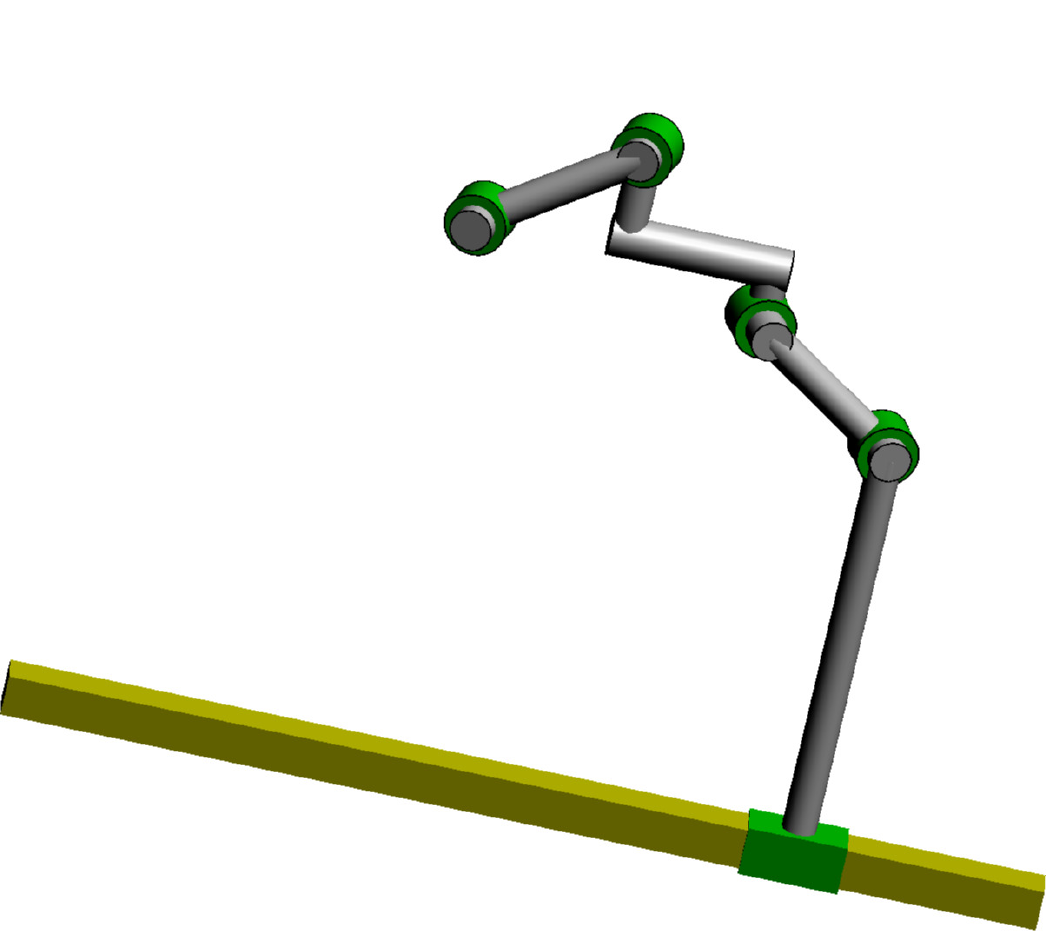}  
  \includegraphics[width=3.04cm]{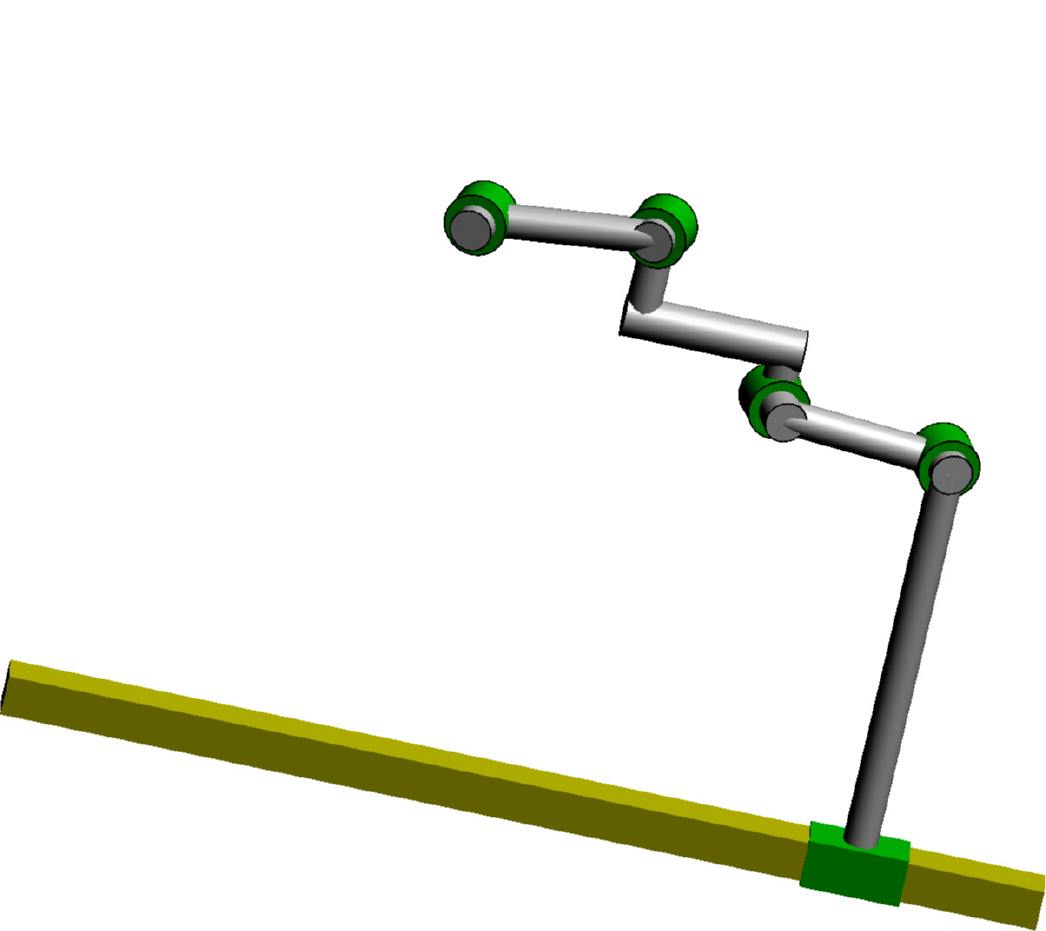}
  \caption{Five configurations of a $P4R$-linkage.}
  \label{fig:p4r}
\end{figure}

We can see these parallel properties from the following example (see \cref{fig:p4r}) which is a concrete known mobile $P4R$-linkage taken from \cite{mavroidis1997spatial,mavroidis1995new}. 
\begin{Ex} \label{ex:p4r}
This is an example of a $P4R$-linkage with two pairs of parallel neighbouring axes.  It is constructed using the geometry constraints from \cite{mavroidis1997spatial,mavroidis1995new}. Five configurations of the  linkage are shown in \cref{fig:p4r}. Set
\begin{equation*}
   h_0 = \eps \qj + \eps \qk, \quad   h_1 = \qk + \eps \qj, \quad
   h_2 = \qk + \eps \qi, \quad
   h_3 = \qj + \eps \qi + 2 \eps \qk,\quad
   h_4 = \qj + \eps \qk.
\end{equation*}
Both sides of 
\begin{equation*}
 (t_0-h_0)(t_1-h_1)(t_2-h_2)(t_3-h_3)(t_4-h_4) \equiv 1
\end{equation*}
are written as 8-dimensional vectors, which yields a system of 7 equations.  The
5th coordinate is a redundant condition because it is already satisfied when the other six
equations are fulfilled due to the Study condition \cite{husty10}. In order to exclude
 ``unwanted'' \ solutions, that is, those such that \[t_{0}(t_{1}^{2}+1)(t_{2}^{2}+1)(t_{3}^{2}+1)(t_{4}^{2}+1) = 0\] in general we add an extra equation 
\[
t_{0}(t_{1}^{2}+1)(t_{2}^{2}+1)(t_{3}^{2}+1)(t_{4}^{2}+1)u-1=0,
\]
Using Gr\"{o}bner basis on Maple, an elimination ideal with respect to $u$ (the Rabinowitsch trick, \cite{Rabinowitsch1929}) is \[I=\langle t_1 - t_3, t_1 + t_2, t_1 + t_4, t_0 t_1 + t_1^2 + t_0 + 1 \rangle\]
which defines the closure of the configuration set $\overline{K_L}$. The set of bonds are then defined by adding the equation $t_{0}(t_{1}^{2}+1)(t_{2}^{2}+1)(t_{3}^{2}+1)(t_{4}^{2}+1)=0$. They are 
\[\{t_0 = 0, t_1 = i, t_2 = -i, t_3 = i, t_4 = -i\} \]
and its conjugate, where $i$ is the complex imaginary unit.
\end{Ex}

\section{\texorpdfstring{$n$}{TEXT}-linkages with mobility at least \texorpdfstring{$n-4$}{TEXT}}\label{newsec:mob2}

In this section we explore the geometric constraints that arise from requiring higher mobility in a linkage, i.e., mobility at least $n-4$ for an $n$-linkage. When we freeze one joint of a linkage $L$ of mobility $m$, we always do it at a generic position. In this case, the mobility of the linkage reduces by 1 since we are adding one equation not in the ideal defining $K_L$. Equivalently, we have a new linkage $L'$ whose mobility is $m-1$.

The structure of this section is as follows: We begin by looking at the case of $6R$-linkages in \cref{6r2}. This is \cref{classificationthm:6rm2}. We then analyse  $6$-linkages which include $P$ and $H$-joints in \cref{6ph}. These are \cref{p6result} and \cref{h6result}, respectively.  We extend this analysis to $n$-linkages where $n>6$ in \cref{nR4} and \cref{n4}. For $nR$-linkages see \cref{rnresult}, for $n$-linkages with $P$-joints \cref{pnresult} and for $n$-linkages with $H$-joints \cref{hnresult}.

%When we say  freeze one joint, we mean that freezing a joint at a generic position and the mobility of the new linkage reduces by 1. 

\begin{Prop} \label{prop:maxmob}
The maximum possible mobility of an $n$-linkage where $n\geq 4$ is $n-3$.
\end{Prop}

\begin{proof}
By the classification of 4-linkages, see \cite{Delassus1922chaines}, this is well known. We proceed by strong induction on $n$. Let $L$ be a linkage with mobility $m$ and $n$ joints, with $m\geq n-2$. If we freeze one joint, then we get a linkage $L'$ of mobility $m-1$ and $n-l$ active joints, where $l\geq 1$. This is a contradiction by the induction hypothesis.
\end{proof}

%\begin{proof}
%By the classification of 4-linkages, see \cite{Delassus1922chaines}, this is well known. We proceed by induction on the mobility.
% Let $L$ be a linkage with mobility $m$ and $n$ joints and assume that $n\leq m+2$. If we freeze one joint, then we get a linkage $L'$ of mobility $m-1$ and $n-l$ active joints, where $l\geq 1$. This is a contradiction by the induction hypothesis.
%\end{proof}

\begin{Prop} \label{prop:SE2}
A closed $n$-linkage with $n\geq 5$ and P or R joints has mobility $n-3$ if and only if the motions of all joints are in one of the two three dimensional groups, $SE(2)$ and $SO(3)$.
\end{Prop}

\begin{proof}
We prove the result by induction on $n$. If $n=5$, the result is well known, see \cite{Ahmadinezhad15}.  We proceed by strong induction on $n$. Let $L$ be a linkage with mobility $m$ and $n$ joints, with $m= n-3$. If we freeze one joint, then we get a linkage $L'$ of mobility $m-1$ and $n-l$ active joints, where $l\geq 1$. By the induction hypothesis and \cref{prop:maxmob}, it follows that $l=1$. Moreover, all the motions of joints of $L'$ are in $SE(2)$ and $SO(3)$. The motion of the joint we froze must also be in the same group due to the closure condition.
\end{proof}

\begin{Rem}
For $n=4$, the statement is not true by several examples: The Bennett 4R-linkage is one of them. %Also, we can not include linkages with helical joints since there are many counter-examples. See \cref{ex:hel}.
\end{Rem}

\begin{Ex} \label{ex:hel}
Consider an $nR$-linkage, $L$, where all revolute axes are parallel. All the joint motions are in $SE(2)$, therefore the mobility of $L$ is $n-3$. Let $L'$ be a new linkage formed by replacing the $R$-joints by $H$-joints with the same pitch. We claim that there is an isomorphism of configurations of $L$ and $L'$, therefore the mobility is the same. If we have a configuration of $L'$, then we can ignore all the translations perpendicular to the plane where all the motions are taking place and this gives a configuration of $L$. In the other direction, we add translations perpendicular to this plane, proportional to the rotation angles. A priori, it is not clear that these translations add up to zero. But they do, since the sum of all rotation angles in $L$ is constant.
\end{Ex}
 %For instance, let $L$ be an $n$R-linkage of mobility $n-3$, then all axes are concurrent (including all axes are parallel). Let $L$ be an $n$-linkage of mobility $n-3$ with at least one P-joint and others are R-joints, then all axes of R-joints are parallel and the direction of P-joints are parallel to the plane which is perpendicular to the axes of R-joints. 
 %Let $L$ be an $n$-linkage of mobility $n-3$ with at least one H-joint and others are R-joints and (or) P-joints, then all axes of R and H-joints are parallel and the direction of P-joints are parallel to the plane which is perpendicular to the axes of R and H-joints. 
 
 A closed linkage has mobility at least $n-4$ if the motions of all joints are in a  Sch{\"o}nflies subgroup of  $SE(3)$ which has dimension 4, see \cite{Herve1978,Herve1999lie,Ravani03}. However, the converse is not true. For instance, the Goldberg 5R is one counter example.

 %We completely classify higher mobility $n$-linkages with mobility $n-4$. The 5-linkages with mobility 1 are classified in \cite{Ahmadinezhad15}. We give the classification for $6R$-linkages with mobility 2 in \cref{6r2}. We give the classification for 6-linkages with mobility 2, $P$-joints and $H$-joints  in \cref{6ph} containing families that are not in a Sch{\"o}nflies subgroup. We give the classification for $nR$-linkages ($n>6$) with mobility $n-4$ or higher in \cref{nR4} where there is no an $nR$-linkage with mobility $n-4$. We give the classification for $n$-linkages ($n>6$) with mobility $n-4$ or higher in \cref{n4} where the $n$-linkages has mobility $n-4$ iff their joints are in a  Sch{\"o}nflies subgroup.

%It turns out that one can derive very concrete conditions on the geometry of the joints of a higher mobility linkage.

Recall that if $L$ has two neighbouring $R$-joints with equal axes or two neighbouring $P$-joints with equal directions,
then we say that $L$ is {\em degenerate}. We also assume that $L$ is not degenerate. No joint parameters are constant during the linkage's motion (the highest dimensional components of the configuration space) (otherwise, one could easily make $n$ smaller). We say that a $P$-joint is {\em perpendicular} to an $R$-joint if the direction of the $P$-joint is perpendicular to the orientation vectors of the $R$-joint. 
% Throughout this section, we assume that $n=5$ or $n=4$, and $L=(j_1,\dots,j_n)$ is a mobile linkage 
% with configuration set $K$.

Throughout this section, we use the following fact when we freeze one joint of an $n$-linkage with mobility $n-4$.
\begin{Lem}\label{n4fr}
Let $L$ be an $n$-linkage, where $n>5$, of mobility $n-4$. If we freeze one joint of $L$, then there is at most one more joint being frozen simultaneously in the resulting linkage $L'$.
\end{Lem}
\begin{proof}
 Recall that we freeze a joint of $L$ at a generic position. If there are other two or more joints being frozen simultaneously, then the resulting linkage $L'$ has $j\leq n-3$ joints with mobility $n-5$ which is impossible.
\end{proof}

\subsection{\texorpdfstring{$6R$}{TEXT}-linkages}\label{6r2}

In this section, we classify paradoxical $6R$-linkages with mobility at least $2$. It is known that all axes of a $6R$-linkage with mobility at least $3$ are concurrent: This is a spherical linkage or a planar linkage. If a $5R$-linkage has mobility $2$, then, by \cref{prop:SE2}, it is either spherical or planar. If a $4R$-linkage has mobility $1$, then, by \cite{Delassus1922chaines}, it can be spherical, planar or Bennett. For a Bennett $4R$-linkage, its Denavit-Hartenberg parameters fulfill: 
 \begin{align} \label{bdh}
\begin{split}
 b_1 & =b_2=b_3=b_4, \\
c_1 & =c_3,\ c_2=c_4, \\
 o_1 & =o_2=o_3=o_4=0.
\end{split}
\end{align}
If a $5R$-linkage has mobility $1$, then, by \cite{Karger1998classification,Hegedues13b},  it is a Goldberg $5R$-linkage.
For a Goldberg $5R$-linkage, its Denavit-Hartenberg parameters fulfill: 
\begin{align}\label{gdh}
 \begin{split}
 b_1 & =b_2=b_3=b_4, \\
c_1 & =c_4, \\
o_2 & =o_3=o_4=0,
\end{split}
\end{align}
and some further complicated equational conditions which are not used in this paper, see \cite{Dietmaier95}. We give the classification of paradoxical $6R$-linkage of mobility $2$ in the following lemma. 

\begin{Lem} \label{lem:aux6rm2}
Let $L=(j_1,j_2,j_3,j_4,j_5,j_6)$ be  a $6R$-linkage with mobility $2$. Suppose we freeze joint $j_1$ at a generic position, and $j_2$ gets frozen as a consequence. Then the relative position of $j_3$ and $j_6$ does not change and the resulting linkage is a Bennett $4R$-linkage.
\end{Lem}

\begin{proof}
Let $L'$ be the linkage resulting from freezing joint $j_1$. We move $L$ and freeze $j_1$ in another position. Again, $j_2$ gets frozen and we call the resulting $4R$-linkage $L''$. Then, $L'$ and $L''$ have the same Denavit-Hartenberg parameters since they are both mobile $4R$-linkages with partially coinciding parameters. If $L'$ is not a Bennett linkage, then $j_3,\,j_4,\,j_5,\,j_6$ are concurrent. 

\textbf{Case I: $j_3,\,j_4,\,j_5,\,j_6$ intersect at a common point.} Since we can replace these four joints by a spherical joint (S), the resulting linkage, denoted by $\tilde{L}$ is a $2RS$-linkage with mobility at least 1 because, otherwise, joints $j_1$ and $j_2$ would not move.  Call the center of the spherical joint $o$. We can compute the orbit of $o$ in the frame of the link with axes $j_1$ and $j_2$. If $o$ is not on the joint $j_1$, then the orbit is a circle around $j_1$ and similarly for $j_2$. These two circles must coincide, otherwise the $2RS$-linkage would not be mobile and we conclude that ${L}$ is a degenerate linkage. On the other hand, if $o$ is on $j_1$, then all axes are concurrent or one joint does not move.

\textbf{Case II: $j_3,\,j_4,\,j_5,\,j_6$ are parallel.} By spherical projection, $j_1$ and $j_2$ are also parallel. Suppose $j_1$ and $j_3$ are not parallel. In that case, if we freeze joint $j_4$ of $L$, the resulting linkage has two pairs/triples of parallel axes which is not possible due to the classification of $4R$ and $5R$-linkages. Therefore, $j_1$ and $j_3$ are parallel. Moreover, the mobility of $L$ needs to be three which contradicts our assumption.
\end{proof}
 
\begin{figure}[tbhp]
  \centering
  \includegraphics[width=3.04cm]{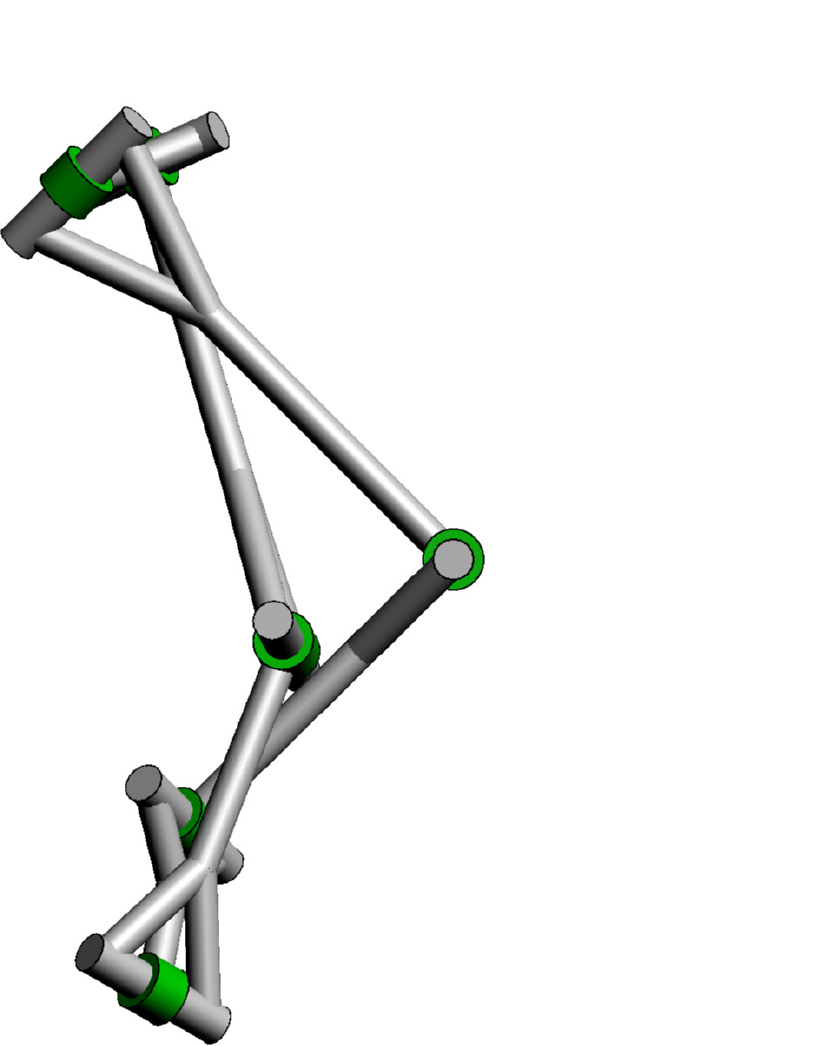}
  \includegraphics[width=3.03cm]{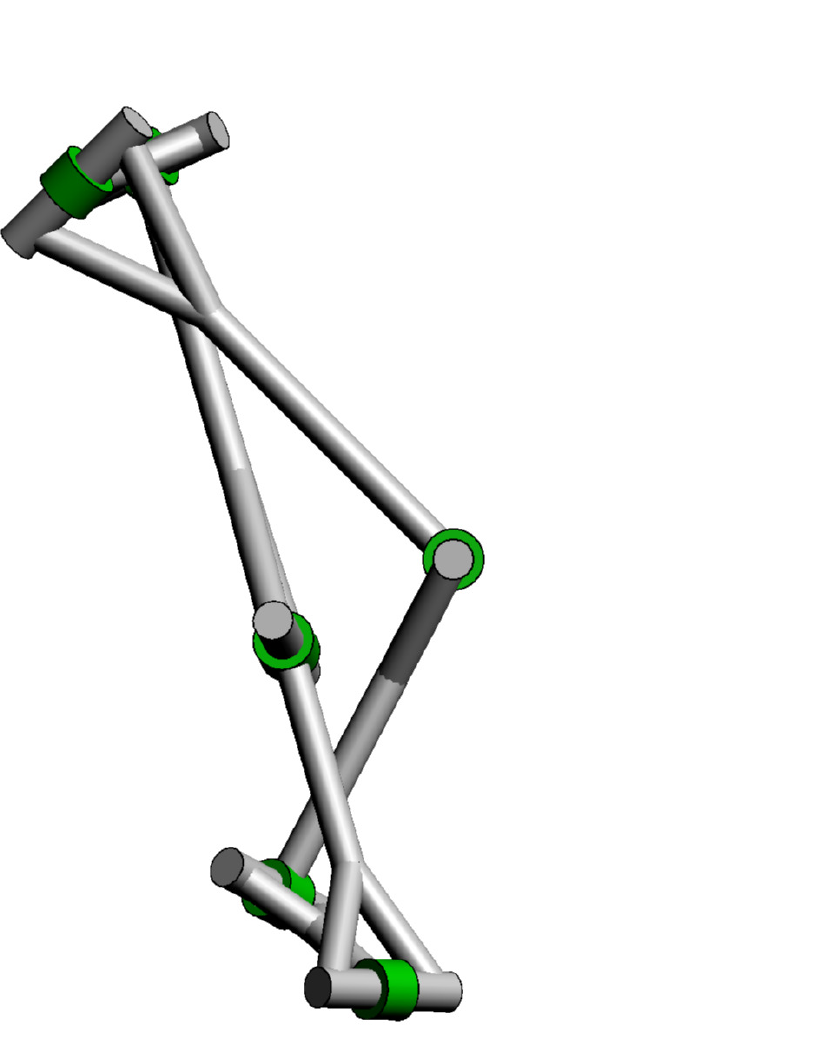}
  \includegraphics[width=3.03cm]{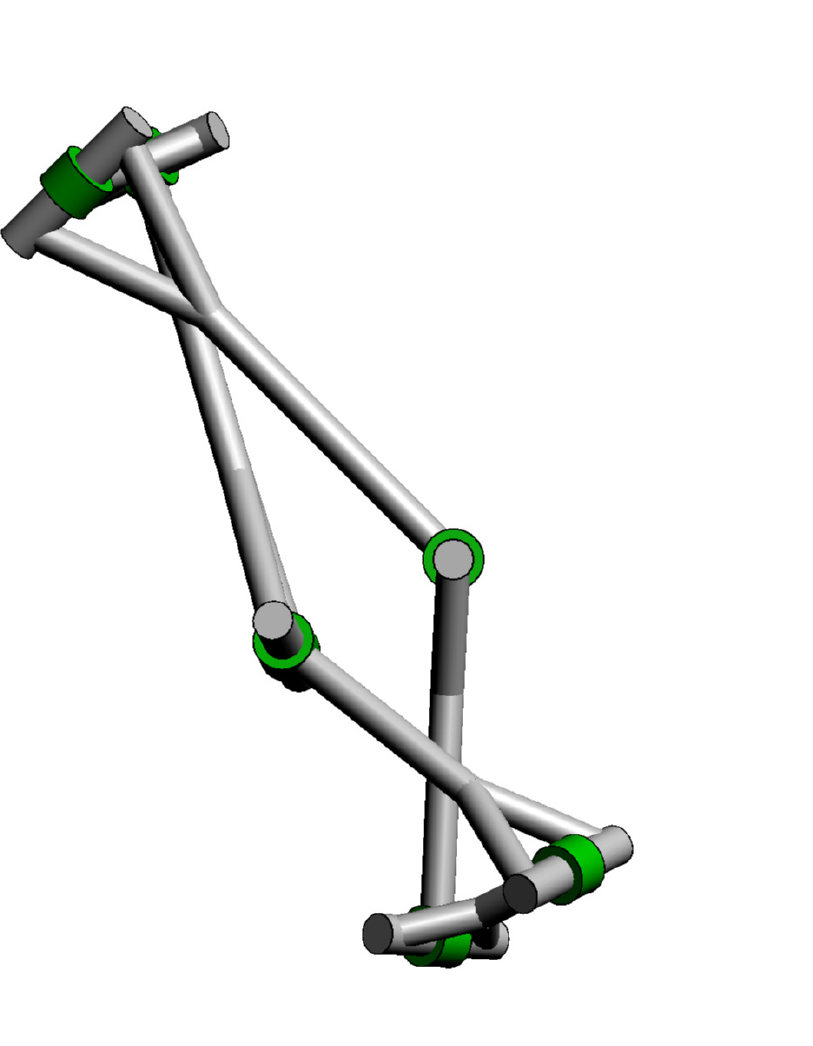}
  \includegraphics[width=3.03cm]{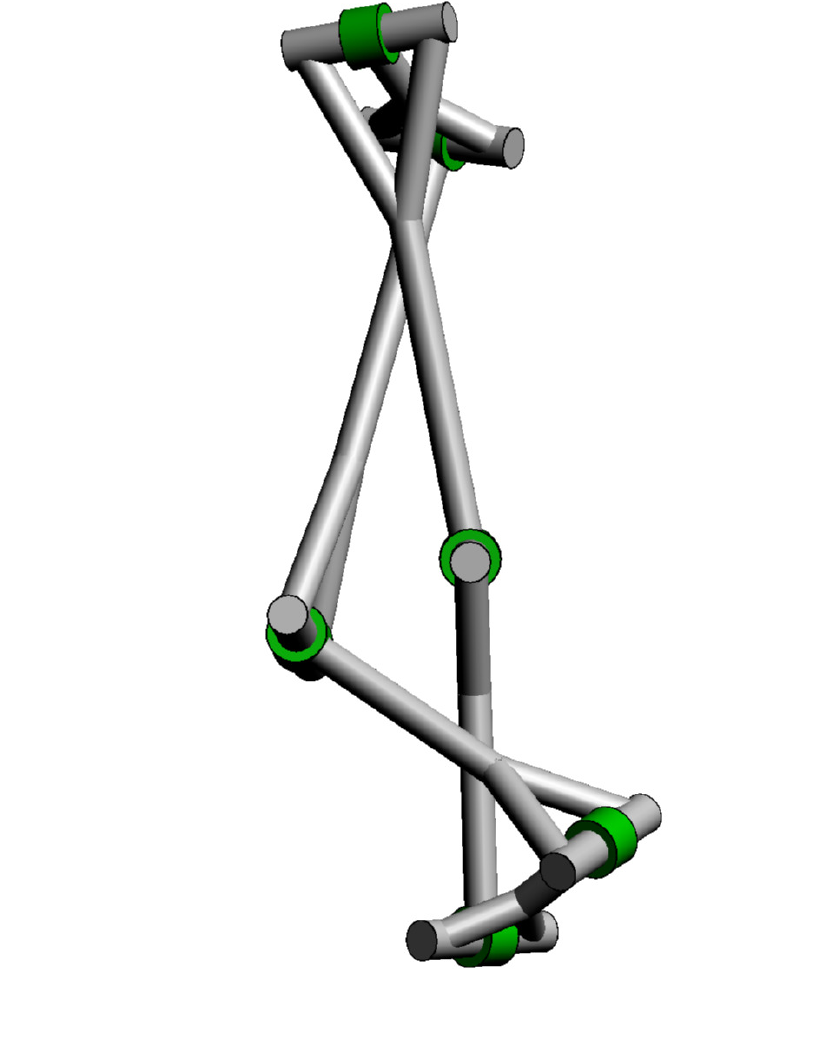}  
  \includegraphics[width=3.04cm]{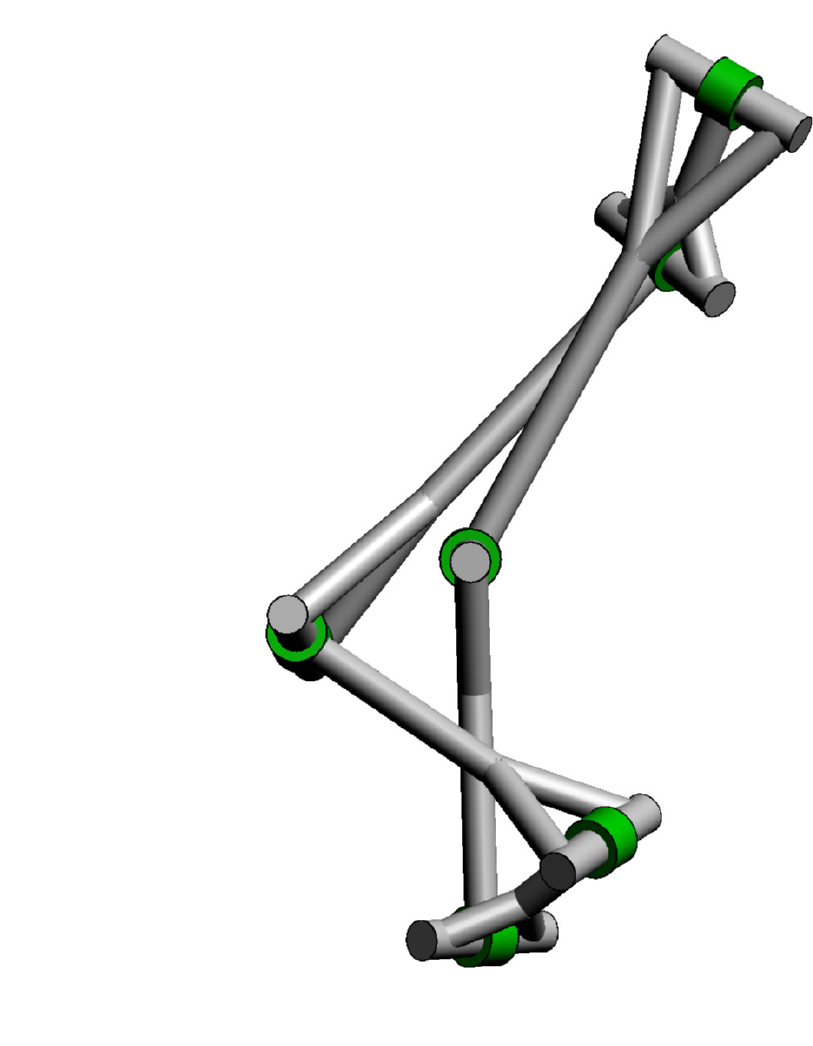}
  \caption{Five configurations of the $6R$-linkage in \cref{ex:6r2-1}.}
  \label{fig:6r2}
\end{figure}

\begin{Thm}\label{classificationthm:6rm2}
Let $L=(j_1,j_2,j_3,j_4,j_5,j_6)$ be  a $6R$-linkage with mobility $2$. Then its Denavit-Hartenberg parameters are: all Bennett ratios are the same, i.e., $b_s=b_{s+1}$, all offsets are zero, i.e., $o_s=0$, and the cosines 
 $c_s$ of twist angles $\alpha_s$ fulfill  one of the following two cases (by a cyclic shift of indices):
\begin{enumerate}
  \item $c_1=c_4,\ c_2=c_6,\ c_3=c_5$. This is a composition of two Bennett Linkages and one common link and two joints.
	\item $c_1=c_4,\ c_2=c_5,\ c_3=c_6$. 
\end{enumerate}
\end{Thm}
\begin{proof}
Fix a 2-dimensional irreducible component of $K_L$. Suppose we freeze one joint. Since $L$ has mobility 2, we still have a mobile linkage $L'$. The number of joints in $L'$ is either $4$ or $5$ by \cref{n4fr}.

\textbf{Case I: $L'$ has 4 mobile joints}. We can assume that the other frozen joint is $j_1$. As the resulting linkage $L'$ is a $4R$-linkage, $L'$ must be a Bennett linkage or a linkage with four concurrent axes. It is not possible that another non-neighbour joint gets frozen, because, by \cref{lem:fr}, all six axes must be concurrent which contradicts the fact that the mobility is 2. Therefore, it is only possible that one of the neighbours of $j_1$ is frozen and we can assume it is joint $j_2$. Then, by \cref{lem:aux6rm2}, the relative position of axes of joints  $j_3$ and $j_6$ does not change when we move $L$ and $L'$ is a Bennett linkage. Therefore we can introduce an extra link connecting $j_3$ and $j_6$ and the linkage $L$ is the composition of two Bennett linkages stacked on top of each other, having one link and two joints in common. See \cref{fig:6r2} for an example.

\textbf{Case II: $L'$ has 5 mobile joints.} The linkage $L'$ becomes a Goldberg $5R$-linkage. The equational conditions of case (2) can be obtained by cyclically freezing joint by joint using the Denavit-Hartenberg parameter constraints in \cref{gdh}. 
\end{proof}

We show the following numeric example of a $6R$-linkage of mobility 2.

\begin{Ex} \label{ex:6r2-1}
Here is an example of a $6R$-linkage without any parallel neighbouring axes with mobility 2. It is constructed using factorization of motion polynomials as in \cite{Hegedues13f}. In fact, it is a combination of two Bennett $4R$-linkages which have two consecutive rotational joints in common. Set
\begin{align*}
   h_1 &= \qi, \\
   h_2 &= \left(\frac{1}{3} + \frac{4}{9}\eps \right)\qi+ 
    \left(\frac{2}{3} + \frac{2}{9}\eps\right)\qj -
    \left(\frac{2}{3} - \frac{4}{9}\eps\right)\qk,\\
   h_3 &= \left(\frac{41}{105}  + \frac{4288}{11025}\eps\right)\qi + 
   \left(\frac{88}{105}  - \frac{16}{11025}\eps\right)\qj  - 
   \left(\frac{8}{21} - \frac{872}{2205}\eps\right)\qk, \\
   h_4 &= \left(\frac{33}{35}  + \frac{68}{1225}\eps\right)\qi -
   \left(\frac{6}{35}  -  \frac{274}{1225}\eps\right)\qj  - 
   \left(\frac{2}{7} - \frac{12}{245}\eps\right)\qk, \\
   h_5 &= \left(\frac{1093}{1365}\eps + \frac{313072}{1863225} \right) \qi - 
   \left( \frac{52}{105}\eps + \frac{16}{11025} \right)\qj + 
   \left(\frac{92}{273} \eps - \frac{149572}{372645} \right) \qk,\\
   h_6 &= \left(\frac{29}{39}+\frac{340}{1521}\eps \right)\qi - 
   \left( \frac{2}{3} - \frac{2}{9}\eps\right)\qj + 
   \left(\frac{2}{39}- \frac{536}{1521}\eps \right)\qk.
\end{align*}
We have that the four joints defined by $h_1, h_2, h_3, h_4$ are a Bennett $4R$-linkage. At the same configuration the $h_1, h_5, h_6, h_4$ are another Bennett $4R$-linkage \cite{Hegedues13f}. 
 Using Gr\"{o}bner basis from Maple, an elimination ideal for the Zariski  closure $\overline{K_L}$ (See \cite{Li15q}) is found to be
 \begin{align*}
  I=\langle 2 t_2 + 2 t_3 + 1, t_6 + t_5 + 1, t_1 t_3 + t_1 t_5 + t_3 t_5 - 1,  \\
  2 t_1 t_3 + 2 t_1 t_5 + 2 t_3 t_4 + 2 t_4 t_5 - 2 t_3 + 3 t_4 - t_5 - 1, \\
   2 t_1 t_3^2 + 2 t_3^2 t_4 - t_1 t_3 + 3 t_1 t_4 - 2 t_3^2 + 3 t_3 t_4 + t_1 - t_3 + 2 t_4 - 1 \rangle.
 \end{align*}
 By adding the equation $(t_{1}^{2}+1)(t_{2}^{2}+1)(t_{3}^{2}+1)(t_{4}^{2}+1)(t_{5}^{2}+1)(t_{6}^{2}+1)=0$ and using  primary decomposition, the 1-dimensional bonds are defined by the ideals:
 \begin{align*}
  B_1 &=\langle t_1 t_3 - t_1 t_6 - t_1 - t_3 t_6 - t_3 - 1,
        t_2 + t_3 + 1/2, 
        t_4 + t_6, 
        t_5 + t_6 + 1,
        t_6^2 + 1 \rangle,\\
  B_2 &=\langle 
        t_1 t_6^2 + 3 t_1 t_6 + 13/4 t_1 + t_4 t_6^2 + 2 t_4 t_6 + 2 t_4 - 1/2 t_6^2 -
            5/4 t_6 - 1/4,
        t_2 + t_4, \\ &\quad\quad
        t_1 t_4 - t_1 t_6 - 3/2 t_1 - t_4 t_6 - t_4 + 1/2 t_6 - 1/2, 
        t_3 - t_4 + 1/2,
        t_4^2 + 1,
        t_5 + t_6 + 1 \rangle,\\
  B_3 &=\langle t_1 + 1/3 t_6 - 1/3,
        t_2 + t_6,
        t_3 - t_6 + 1/2,
        t_5 + t_6 + 1,
        t_6^2 + 1 \rangle,\\
  B_4 &=\langle t_1 + t_3,
        t_2 + t_3 + 1/2,
        t_3^2 + 1,
        t_3 t_4 + t_3 t_6 - t_4 t_6 + 1/2 t_4 + 1/2 t_6 + 1, \\ &\quad\quad
        t_3 t_6^2 + t_3 - t_4 t_6^2 + t_4 t_6 - 5/4 t_4 + 1/2 t_6^2 - 1/4 t_6 - 1/2, \\  &\quad\quad
        t_4^2 t_6^2 - t_4^2 t_6 + 5/4 t_4^2 - t_4 t_6^2 + 1/2 t_4 t_6 + t_4 + 5/4 t_6^2 +
            t_6 + 1,
        t_5 + t_6 + 1 \rangle,\\
  B_5 &=\langle t_1 - t_6 - 1,
        t_2 + t_3 + 1/2,
        t_3 t_4 + t_3 t_6 - t_4 t_6 + 1/2 t_4 + 1/2 t_6 + 1,  \\&\quad\quad
        t_5 + t_6 + 1,
        t_6^2 + 2 t_6 + 2 \rangle,\\
  B_6 &=\langle t_2 + t_6 + 3/2,
        t_3 - t_6 - 1,
        t_4 - 1/3 t_6 - 2/3, 
        t_5 + t_6 + 1,
        t_6^2 + 2 t_6 + 2 \rangle.
 \end{align*}
Five configurations of the mobility 2 linkage are shown in \cref{fig:6r2}.
\end{Ex}

\begin{Ex} \label{ex:6r2-2}
There is a cheap way to obtain another linkage when the original linkage has a Bennett triple. 
%satisfies a Bennett condition. 
 This is called the isomeric mechanism by \cite{wohlhart1991isomeric}. Namely, one can replace a joint axis (which is  the middle one of a Bennett triple) by another new axis where the new axis and the original three axis make a Bennett loop.  There are six consecutive Bennett triples in \cref{ex:6r2-1}. Using the isomerization for some consecutive Bennett triples, we will obtain $6R$-linkages with frozen joints which are not mainly interesting here. One can obtain other three isomeric mechanisms by replacing  one (or two) of the joints $h_1$ and $h_4$ by new joint axes $h'_1$ and $h'_4$
\begin{align*}
   h'_1 &= \left(\frac{101}{117} + \frac{112}{1521}\eps \right)\qi+ 
    \frac{4}{9}\qj - \left(\frac{28}{117} - \frac{404}{1521}\eps\right)\qk,\\
   h'_4 &= \left(\frac{3301}{4095}  + \frac{240628}{1863225}\eps\right)\qi +
   \left(\frac{86}{315}  +  \frac{274}{1225}\eps\right)\qj  - 
   \left(\frac{430}{819} - \frac{117232}{372645}\eps\right)\qk. 
\end{align*}
With this replacement, we have no Bennett $4R$-linkage for any consecutive four joints. 
 %Using Gr\"{o}bner basis from Maple, we can find the Zariski closure of the configuration space of this linkage, similarly to example \cref{ex:6r2-1}.
 \end{Ex}

\subsection{\texorpdfstring{$6$}{TEXT}-linkages}\label{6ph}
 
In this section we classify 6-linkages of mobility 2 or higher containing revolute joints and at least one prismatic or helical joint. Mobile linkages with 4 joints of type $R$, $P$ or $H$ have been classified in \cite{Delassus1922chaines}. Mobile linkages with 5 joints of type $R$, $P$ or $H$ have been classified in \cite{Ahmadinezhad15}.

The following result settles the case for 5-linkages with $P$-joints and $R$-joints:
\begin{Thm}[{\cite[Theorem~6]{Ahmadinezhad15}}]\label{p4r}
Let $L$ be a mobile 5-linkage  with at least one $P$-joint and all other joints of type $R$. Then the following two  cases are possible.
\begin{enumerate}
	\item Up to cyclic shift, $j_0$ is the only $P$-joint, $h_1 \parallel h_2$ and $h_3\parallel h_4$,  and $t_1=\pm t_2$ and $t_3=\pm t_4$ is fulfilled on the configuration curve.
	\item All axes of $R$-joints are parallel.
\end{enumerate}
\end{Thm}

\begin{Rem}\label{p4r:rmk}
 Let $L$ be a 5-linkage with at least one $P$-joint and all other joints are parallel of type $R$. 
 If the mobility of $L$ is 2 then all $P$-joints are perpendicular to the $R$-joints. If the mobility of $L$ is 1 then there are at least two $P$-joints which are not perpendicular to the $R$-joints.
 %Let $\Pi$ be the plane which is perpendicular to the $R$-joints.  %If there is only one $P$-joint and all $R$-joints are parallel, then the mobility of $L$ is 3 provided that the direction of the $P$-joint is parallel to normal plane to the axes of the $R$-joints. 
% If the mobility of $L$ is 2 then the direction of all $P$-joints is parallel to $\Pi$. If the mobility of $L$ is 1 then there are at least two $P$-joints whose directions are not parallel to the plane $\Pi$.
 %If there is just one $P$-joint in $L$, then the mobility of $L$ is 2. If there are two $P$-joints or more in $L$, then the mobility of $L$ is either 2 where the directions of all $P$-joints are parallel to the plane (say $M$) which is perpendicular to the axes $R$-joints or 1 where there are at least two $P$-joints such that their directions are not parallel to the plane $M$.
\end{Rem}

% Any $6-$linkage of mobility $2$ or higher containing at least one prismatic joint has parallel revolute axes.

Complementing \cref{classificationthm:6rm2}, we solve the classification of linkages with 6 joints of mobility $2$ with joints type $R$ and $P$:
\begin{Thm}\label{p6result}
Let $L$ be a 6-linkage of mobility $2$ with at least one $P$-joint and all other joints of type $R$. Then one of two following cases occur:
\begin{enumerate}
	\item All axes of $R$-joints are parallel and $L$ has at least two $P$-joints.
	\item Up to cyclic shift, $j_0$ and $j_3$ are the only two $P$-joints,  $h_1 \parallel h_2$, $h_5\parallel h_4$, $p_0 \parallel p_3$,  and $t_1=\pm t_2$ and $t_5=\pm t_4$ is fulfilled on the configuration curve. %Suppose $j_0$ and $j_3$ are the only two $P$-joints of $L$. Then, $h_1 \parallel h_2$, $h_5\parallel h_4$, $p_0 \parallel p_3$,  and $t_1=\pm t_2$ and $t_5=\pm t_4$ is fulfilled on the configuration curve.
\end{enumerate}
\end{Thm}

\begin{figure}[tbhp]
  \centering
  \includegraphics[width=3.04cm]{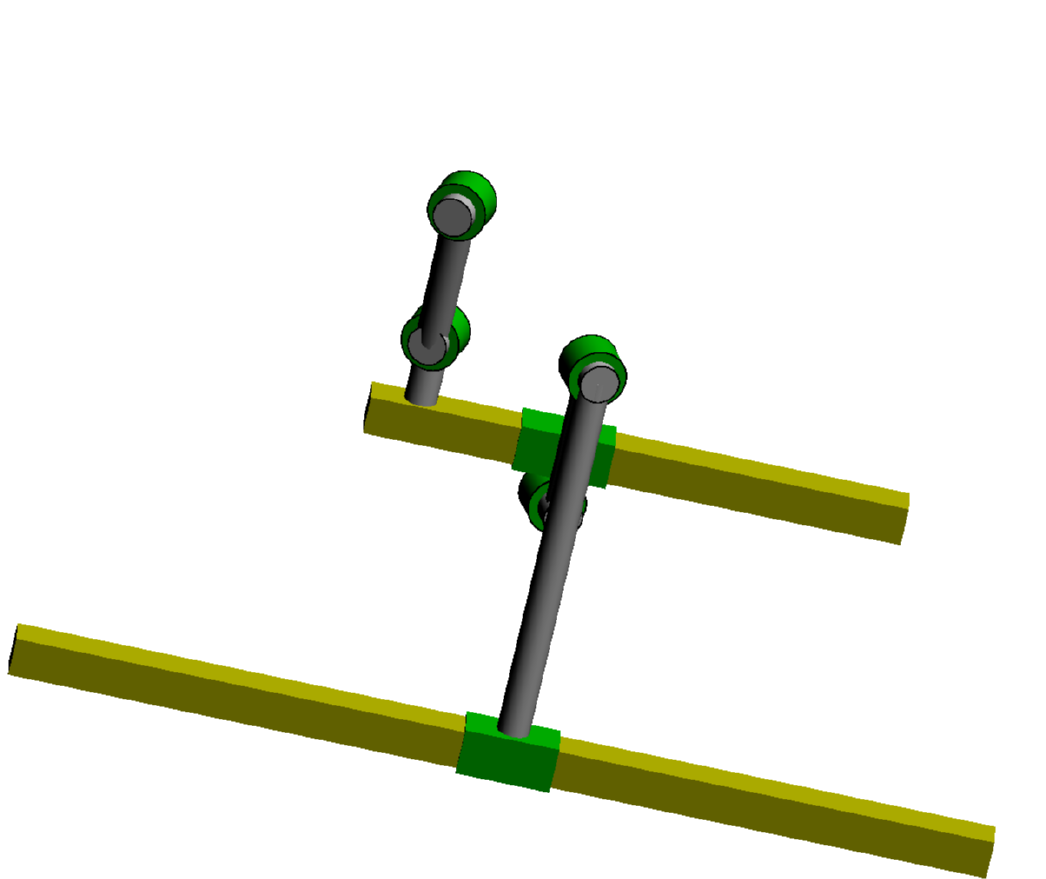}
  \includegraphics[width=3.03cm]{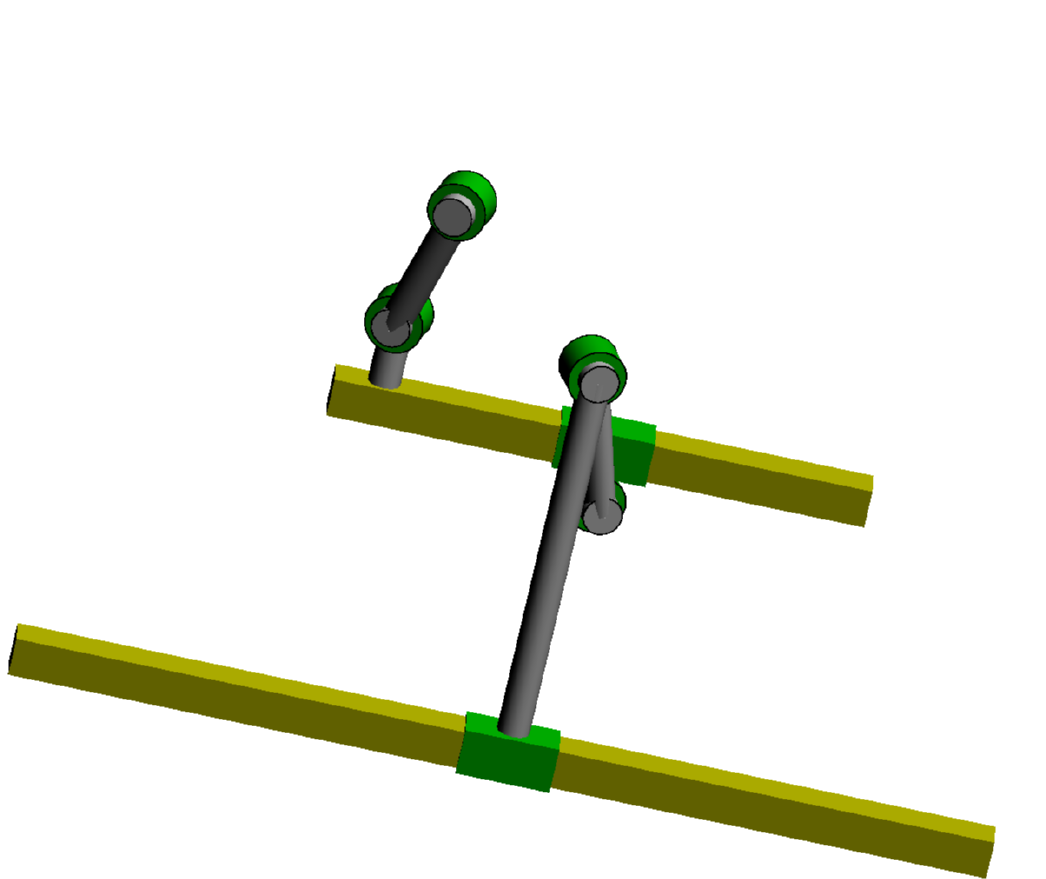}
  \includegraphics[width=3.03cm]{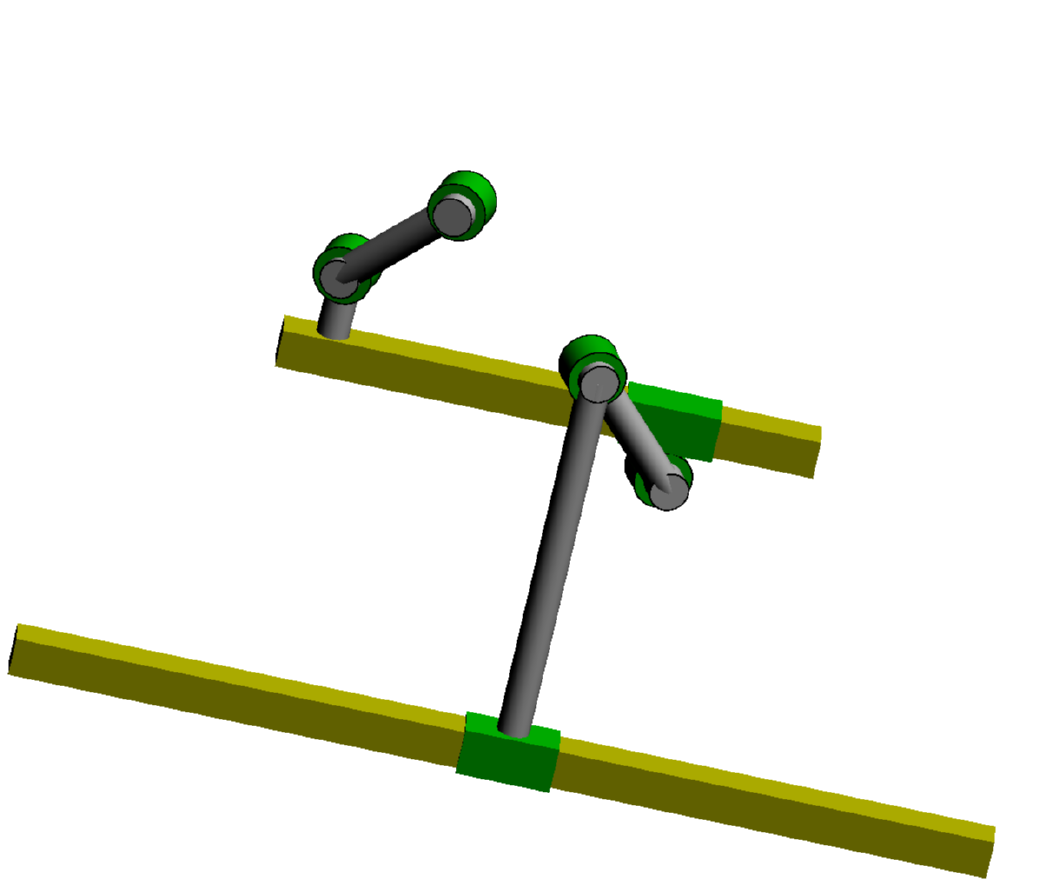}
  \includegraphics[width=3.03cm]{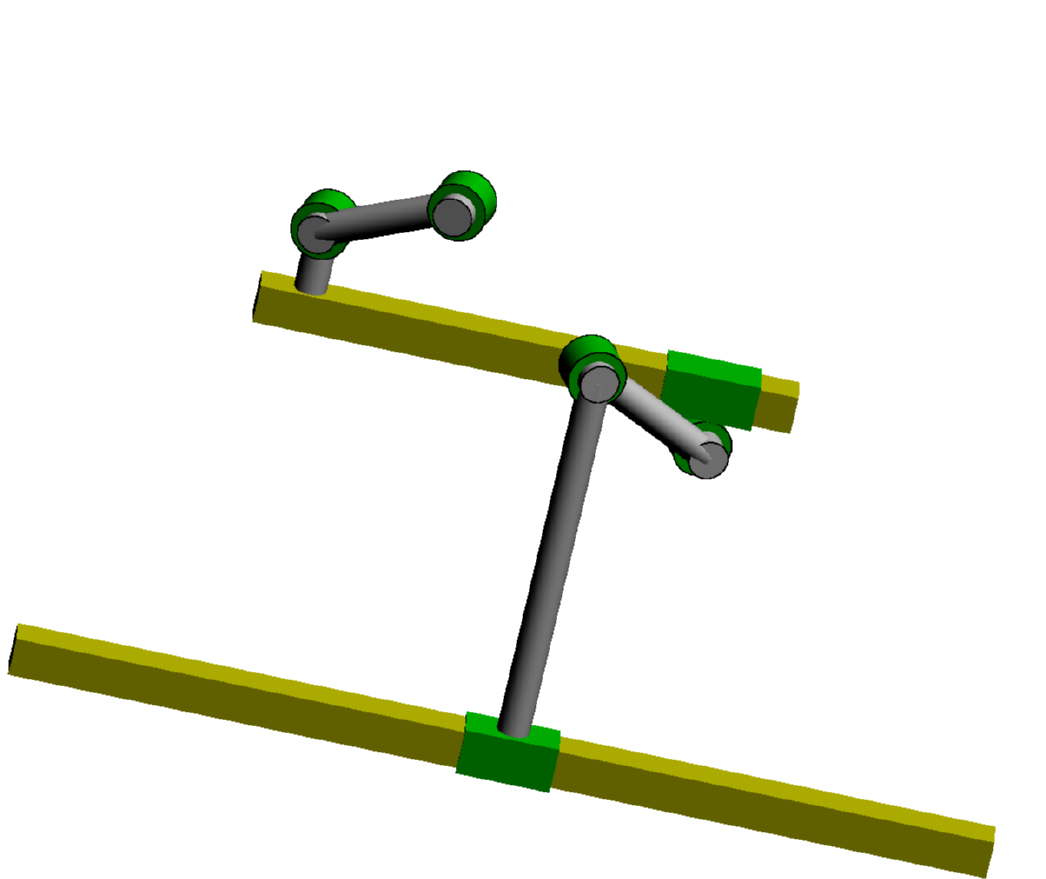}  
  \includegraphics[width=3.04cm]{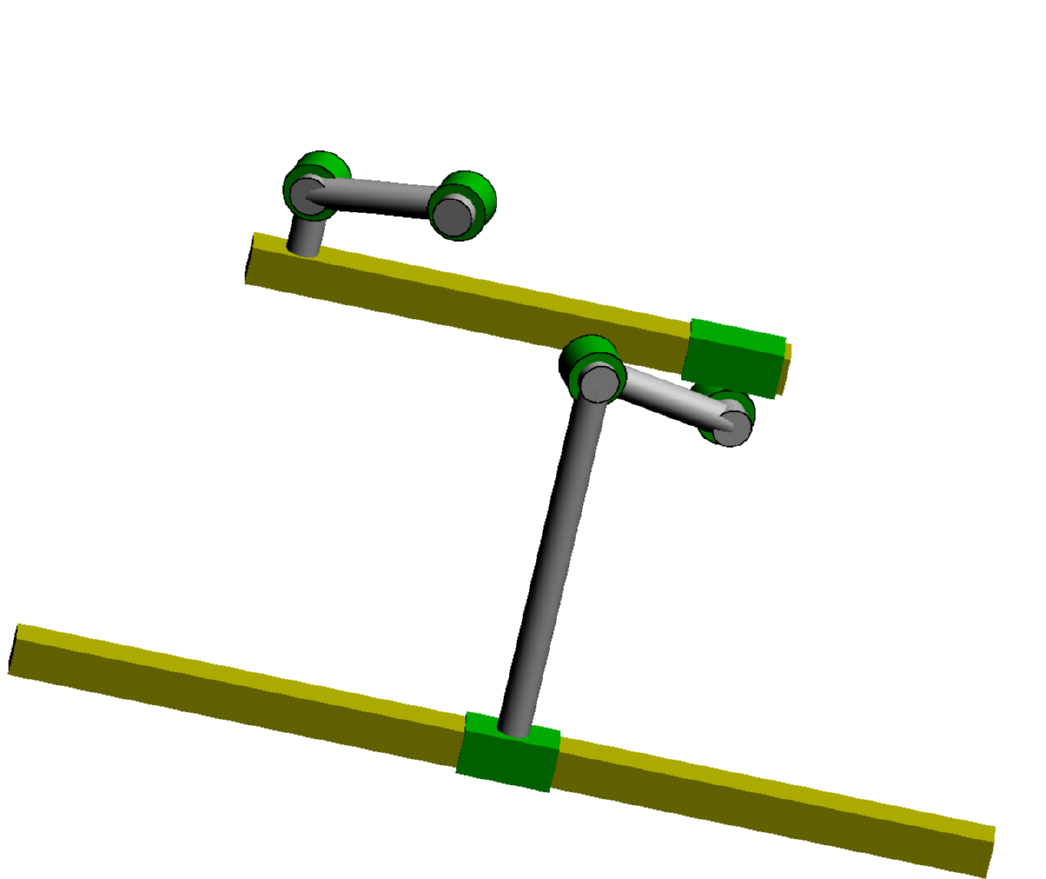}
  \caption{Five configurations of a $PRRPRR$-linkage.}
  \label{fig:prrprr}
\end{figure}

\begin{proof} 
We make a case distinction based on the number of $P$-joints.

\textbf{Case I: $L$ has one $P$-joint}. We claim that this case is not possible. We prove the claim by finding contradictions. Since $L$ has mobility 2, we can find an $R$-joint such that the $P$-joint is still active when we freeze this $R$-joint. If we freeze such an $R$-joint, we still have a mobile linkage $L'$. We now have two possibilities. Either no other joint is frozen, or another $R$-joint gets frozen. If no other joint is frozen, then the resulting linkage is a $P4R$-linkage of mobility 1. Moreover, by the classification in \cref{p4r}, we have two pairs of parallel joints, or all axes are parallel by \cref{p4r} and \cref{p4r:rmk}. 
\begin{itemize}
	\item All axes of $R$-joints are parallel. Then all axes of $R$-joints in $L$ are parallel using the spherical projection. Moreove, $L$ has mobility two if and only if the $P$-joint is frozen because, otherwise, the mobility is three by \cref{prop:SE2} which contradicts the fact that all joints of $L$ are movable.
	\item The $L$ has exactly two pairs of parallel joints. Therefore, we can freeze the $P$-joint of $L$, resulting in a $5R$-linkage with two pairs of parallel joints in different directions which is not possible by the classification of $5R$-linkages \cite{Karger1998classification, Hegedues13b}. 
\end{itemize}
On the other hand, if another $R$-joint is frozen, then the resulting mobile 4-linkage $L'$ will become a 4-linkage with at least one $P$-joint. Therefore, all axes of $R$-joints are parallel in $L'$ by \cite{Delassus1922chaines}. In addition, if we freeze the $P$-joint, all axes of $R$-joints are parallel in $L$. Hence, the $P$-joint must be perpendicular to the  $R$-joints. Then the mobility of $L$ needs to be three, which contradicts our assumption. 

\textbf{Case II: $L$ has two $P$-joints}. We freeze one $P$-joint (say, joint $j_6$) and call the resulting linkage $L'$. If the other $P$-joint is frozen too, then the resulting $4R$-linkage is mobile when all axes are parallel. A Bennett or spherical $4R$-linkage is not possible because freezing a $P$-joint can change the normal feet or twist distances. If the other $P$-joint is active, then we claim that the resulting mobility of one linkage $L'$ has five active joints. In other words, we have two pairs of neighbouring parallel joints in $L'$ by \cref{p4r} and \cref{p4r:rmk}. Therefore, in this case, we have two possibilities, either all axes of $R$-joints are parallel, or there are two pairs of parallel $R$-joints as in \cref{p4r} (see \cref{fig:p4r}). In addition, we know that the $P$-joints are not perpendicular to the $R$-joints in $L'$, and the same holds for $L$. We prove the claim by contradiction. Assume that if we freeze the joint $j_6$, another $R$ joint is frozen simultaneously. Then the resulting $L'$ is a mobile 4-linkage with a $P$-joint by \cref{n4fr}. The $P$-joint is perpendicular to the $R$-joints in $L'$ the by the classification of 4-linkages. The axes of $R$-joints of $L$ must be parallel by spherical projection. In addition, we have the contradictions that the mobility of $L$ is 3 when the joint $j_6$ is active, perpendicular to the $R$-joints, or the joint $j_6$ is not an active joint in $L$. Furthermore, we claim that the two $P$-joints have the same direction when there are two pairs of parallel $R$-joints in $L$ (see \cref{fig:prrprr}). Otherwise, if we freeze an $R$-joint in $L$ and denote the resulting linkage by $L''$, then a neighbouring $R$-joint must be frozen because of two pairs of parallel $R$-joints. Hence, the two remaining $R$-joints must generate a circular translation, and the resulting linkage is a 4-linkage with two $P$-joints. Then the $P$-joints are perpendicular $R$-joints in $L''$, which is a contradiction.

\textbf{Case III: $L$ has $k=3$ or more $P$-joints}. Then all rotation axes are parallel. Because the spherical projection of $L$ with $6-k\le 3$ revolute joints is degenerate: all axes are coinciding. Therefore, the mobility of $L$ is 2 when at least two $P$-joints are not perpendicular to the $R$-joints.
\end{proof}

\begin{Rem}\label{6p:rmk}
Let $L$ be a 6-linkage of mobility 2 or 3 with at least one $P$-joint and all other joints are parallel $R$-joints. 
If the mobility of $L$ is 3 then all $P$-joints are perpendicular to the $R$-joints. If the mobility of $L$ is 2 then there are at least two $P$-joints which are not perpendicular to the $R$-joints. Compare with \cref{p4r:rmk}.
%  Let $L$ be a 6-linkage of mobility 2 or 3 with at least one $P$-joint and all other joints are parallel $R$-joints. 
% Let $\Pi$ be the plane which is perpendicular to the $R$-joints.  %If there is only one $P$-joint and all $R$-joints are parallel, then the mobility of $L$ is 3 provided that the direction of the $P$-joint is parallel to normal plane to the axes of the $R$-joints.
% If the mobility of $L$ is 3 then the direction of all $P$-joints is parallel to $\Pi$. If the mobility of $L$ is 2 then there are at least two $P$-joints whose directions are not parallel to the plane $\Pi$. Compare with \cref{p4r:rmk}.
%If there is just one $P$-joint in $L$, then the mobility of $L$ is 3 when the direction of the $P$-joint is parallel to the plane which is perpendicular to the axes of $R$-joints when these are all parallel. 
%If there are two $P$-joints or more in $L$ and the axes of $R$-joints are parallel, then the mobility of $L$ is either 3 or 2. It is 3 when the directions of all $P$-joints are parallel to the plane, say $M$, which is perpendicular to the axes of $R$-joints. It is 2 when there are at least two $P$-joints whose directions are not parallel to the plane $M$.
\end{Rem}

Mobile linkages with 5 joints of type $R$, $P$ or $H$ have been classified in \cite{Ahmadinezhad15}.
\begin{Thm}[{\cite[Theorem~10]{Ahmadinezhad15}}]\label{h5}
Let $L$ be a non-degenerate mobile 5-linkage with $R$-, $P$-, and $H$-joints, with at least one $H$-joint. Up to cyclic permutation, the following cases are possible.
\begin{enumerate}
\item All axes of $R$- and $H$-joints are parallel. 
\item There is one $P$-joint $j_0$, all other joints are of type $H$ or $R$, $h_1||h_2$ and $h_3||h_4$.
\end{enumerate}
\end{Thm}
We solve the case for linkages with 6 joints of mobility $2$ with joints type $R$, $P$ or $H$:
\begin{Thm}\label{h6result}
% Any $6-$linkage of mobility $2$ or higher containing at least one prismatic joint has parallel revolute axes.
Let $L$ be a 6-linkage of mobility 2 or higher with at least one  $H$-joint. Up to cyclic permutation, the following cases are possible.
\begin{enumerate}
\item All axes of $R$- and $H$-joints are parallel. 
\item There are two $P$-joints $j_0$ and $j_3$ and $p_0||p_3$. All other joints are of type $H$ or $R$, $h_1||h_2$ and $h_4||h_5$.
\end{enumerate}
\end{Thm}

\begin{proof}
% It is similar to the proof of \ref{p6result}. We only need to use \ref{h5} instead of \ref{p4r}.
Recall that if two axes of a linkage are parallel then they coincide in the spherical projection. Let $k$ be the number of blocks of neighbouring equal axes of the spherical projection $L_s$. By \cref{rem:Lr}, we know that mobility of $L_r$ is greater than or equal to the mobility of $L$. Then the mobility of $L_r$ is at least two. Furthermore, if the rotation joints of $L_r$ are concurrent, then they must be parallel because at least one $R$-joint comes from an $H$-joint. Therefore, by \cref{classificationthm:6rm2} and \cref{p6result}, there are at most two blocks of equal axes of the spherical projection $L_s$.  
The proof proceeds by case distinction on $k$. 

\begin{description}
\item[Case $k=1$] Then all axes of $L_s$ are equal; hence all axes of $H$- and $R$-joints of $L$ are parallel; this is possibility~(1) of the theorem, and the motion of each joint belongs to the same Sch{\"o}nflies motion subgroup of $SE(3)$.

\item[Case $k=2$] Then each of the two blocks of $R$-joints in $L_s$ has at least two joints. Indeed, if a block has a single joint, by the spherical projection, it must be parallel to other $R$-joints which contradicts our assumption of two blocks. The linkage $L_r$ is movable and, therefore, has at least four joints that move. In particular, it cannot happen that all axes of $L_r$ that move are parallel. After removing the fixed joints, $L_r$ still has two blocks of parallel axes. By comparing with \cref{p6result}, it follows that $L_r$ is a $PRRPRR$-linkage; if, say, $j_0$ and $j_3$ are prismatic joints, then $h_1||h_2$ and $h_4||h_5$ which is the possibility~(2) of the theorem.

% \item[Case $r\ge 3$] If at least one group of joints of the spherical projection $L_s$ has only one $R$-joint, then this joint cannot move. Hence the corresponding $H$- or $R$-joint of $L$ does not move either, contradicting our assumption. Hence no group of joints of the spherical projection $L_s$ with only one $R$-joint. By freezing one joint of $L$, at most, two joints are frozen simultaneously by \cref{n4fr}, say $L'$ with a helical joint is the resulting linkage. The linkage $L'$ cannot be a 5-linkage or a 4-linkage. Indeed, $L'$ has three groups of joints of the spherical projection, which is impossible for a 5-linkage with a helical joint. If $L'$ is a 4-linkage with a helical joint, then $L'$ has at least two groups of joints of the spherical projection, which is impossible too. 
\end{description}\end{proof}

\subsection{\texorpdfstring{$n$R}{TEXT}-linkages}\label{nR4}

In this section, we classify paradoxical $nR$-linkages with mobility of $n-4$, where $n>6$.

\begin{Thm}\label{rnresult}
Let $L$ be an $nR$-linkage, where $n>6$, of mobility $n-4$ or higher, then  all axes are concurrent. Moreover, $L$ has mobility $n-3$.
\end{Thm}
\begin{proof}

We prove this result by induction on the number of active joints.

For $n = 7$, we assume that $L$ is a 7$R$-linkage with mobility $3$. Then, freezing the joint $j_{n}$ of $L$, we get a linkage $L'$ with mobility  $2$. 
\begin{description}
\item[No other joint is frozen] by \cref{classificationthm:6rm2}, the triples $(h_2,h_1,h_{n-1})$ and $(h_1,h_{n-1},h_{n-2})$ satisfy  Bennett conditions and, in fact, all the Bennett ratios are equal. As we freeze the joint $j_n$ at an arbitrary position, we get that $h_2, h_1, h_{n}, h_{n-1}$ are concurrent by \cref{lem:fr}. Therefore, all Bennett ratios are equal to zero, and $L'$ has mobility 3. Then $L$ has mobility 4, which is a contradiction.
\item [Another joint is frozen] $L'$ becomes a $5R$-linkage with mobility at least 2. Then all axes of $L'$ are concurrent. We can freeze another joint of $L$ different from $j_n$ and, in a similar way, we conclude that all axes of the resulting linkage are concurrent. Therefore, all axes are concurrent, and the mobility of $L$ is 4. 
\end{description}

 Fix $n \geq 8$. Suppose that all axes of any $(n-k)R$-linkage with $k\geq 1$ whose mobility is at least $n-5$ are concurrent. Let $L$ be an $nR$-linkage with mobility at least $n-4$. Then, freezing the joint $j_{n}$ of $L$, we get an $(n-k)R$-linkage $L'$ with mobility at least $n-5$. By strong induction hypothesis, $L'$ has mobility $n-4$ and all axes of $L'$ (and hence of $L$) are concurrent.\end{proof}

\subsection{\texorpdfstring{$n$}{TEXT}-linkages}\label{n4}

In this section, we classify paradoxical $n$-linkages with mobility $n-4$, where $n>6$. If the mobility of an $n$-linkage with at least one  $P$-joint and other $R$-joints is $n-3$, then all the revolute joints are parallel. Compare the following lemma with \cref{p6result}.

\begin{Lem} \label{p7result}
Let $L$ be an $7$-linkage of mobility 3  with at least one  $P$-joint and other $R$-joints. Then all axes of $R$-joints are parallel and the number of $P$-joints is at least 2.
\end{Lem}
\begin{proof}
We make a case distinction based on the number of $P$-joints.

\textbf{Case I: $L$ has one $P$-joint}. Since $L$ has mobility 3, we can find an $R$-joint such that the $P$-joint is still active when we freeze this $R$-joint. If we freeze such an $R$-joint, we still have a mobile linkage $L'$ with mobility 2. If no other joint is frozen, then we have a contradiction by \cref{p6result}. On the other hand,  by \cref{n4fr}, if another $R$-joint is frozen, then the resulting mobile 5-linkage $L'$  has mobility 2 with one active $P$-joint. Hence, all axes of $R$-joints need to be parallel in $L'$. If we have just one group of paralell joints in $L$, where the two frozen $R$-joints are parallel to the axes of $R$-joints, i.e., all axes of $R$-joints in $L$ are parallel, or parallel to each other and neighbouring in $L$, which contradicts the assumption. If we have two groups of parallel joints in $L$, we freeze the $P$-joint of $L$. In that case, the resulting linkage with at most six revolute joints has mobility of two and two groups of parallel joints by \cref{n4fr}, which is impossible by the classification of $5R$-linkages and $6R$-linkages with mobility 2, c.f.~\cref{classificationthm:6rm2}. Therefore, the $L$ with one $P$-joint is impossible by above arguments. %Since $L$ has mobility 3, we can find an $R$-joint such that the $P$-joint is still active when we freeze this $R$-joint. If we freeze such an $R$-joint, we still have a mobile linkage $L'$ with mobility 2. If no other joint is frozen, then we have contradiction by \cref{p6result}.  On the other hand,  by \cref{n4fr}, if another $R$-joint is frozen, then the resulting mobile 5-linkage $L'$  has mobility 2 with one $P$-joint. Then all axes of $R$-joints are parallel in $L'$.  The two frozen $R$-joints are either parallel to the axes of $R$-joints, i.e., all axes of $R$-joints in $L$ are parallel, or parallel to each other and being neighbouring in $L$. Assuming that we have two groups of parallel joints in $L$. If we freeze the $P$-joint of $L$, the resulting linkage with at most six revolute joints has mobility 2 and two groups of parallel joints. This is impossible by the classification of $5R$-linkages and $6R$-linkages with mobility 2, c.f.~\cref{classificationthm:6rm2}. Therefore, in this case, the direction of the $P$-joint must be perpendicular to the direction of the axes of $R$-joints, and the mobility of $L$ is 4 which contradicts our assumption.

\textbf{Case II: $L$ has two $P$-joints}. If we freeze a $P$-joint, we still have a mobile linkage $L'$ with mobility 2. If the other $P$-joint is frozen, then the resulting $L'$ is a $5R$-linkage of mobility 2. Then all axes of $R$-joints must be parallel because the $P$-joint would change the intersecting point if all axes of $R$-joints were concurrent. Notice that the $P$-joints are not perpendicular to the $R$-joints. On the other hand, if the other $P$-joint is not frozen, then the resulting linkage must be a 5-linkage by \cref{p6result} and  \cref{prop:SE2} and all axes of $R$-joints are parallel in $L'$. Therefore, the frozen $R$-joint must be parallel to the axes of $R$-joints in $L'$, i.e., all axes of $R$-joints in $L$ are parallel, and the direction of one $P$-joint is parallel to the plane that is perpendicular to the axes of $R$-joints which contradicts that the mobility of $L$ is 3.  %Therefore, in this case, the directions of the two $P$-joints are not parallel to the plane which is perpendicular to the axes of $R$-joints.

\textbf{Case III: $L$ has three $P$-joints}. If we freeze a $P$-joint (named $j_n$), we still have a mobile linkage $L'$ with mobility 2. The other two $P$-joints cannot be frozen simultaneously. Otherwise, the resulting linkage $L'$ is an impossible $4R$-linkage of mobility 2. On the one hand, if another $P$-joint is frozen, then the resulting mobile 5-linkage $L'$  has mobility 2 with one $P$-joint by \cref{n4fr}. Then all axes of $R$-joints are parallel in $L'$, and all axes of $R$-joints must be parallel in $L$. On the other hand, if the two $P$-joints are both active, then either all axes of $R$-joints are parallel in $L'$, or we have two pairs of parallel joints and two $P$-joints with the same direction by \cref{p6result}. Assume that we have only one group of parallel joints, all axes of $R$-joints are parallel in $L$. Assume that we have two pairs of parallel joints, we can freeze another $P$-joint of $L$ different from $j_n$ and, in a similar way, we conclude that the directions of all three $P$-joints are the same and two of them are neighbouring joints by \cref{p6result}. Then $L$ must degenerate, which is a contradiction. Therefore, all axes of $R$-joints are parallel in $L$, and there are at least two $P$-joints such that they are not perpendicular to the $R$-joints. %If we freeze a $P$-joint (which is named $j_n$), we still have a mobile linkage $L'$ with mobility 2. The other two $P$-joints cannot be frozen simultaneously. Otherwise, the resulting linkage $L'$ is a $4R$-linkage of mobility 2 which is impossible. On one hand, if another $P$-joint is  frozen, then the resulting mobile 5-linkage $L'$  has mobility 2 with one $P$-joint. Then all axes of $R$-joints are parallel in $L'$ and all axes of $R$-joints must be parallel in $L$. On the other hand, if the two $P$-joints are both active, then either all axes of $R$-joints are parallel in $L'$ or we have two pairs of parallel joints and two $P$-joints with the same direction by \cref{p6result}. Assuming we have only one group of parallel joints, then  all axes of $R$-joints must be parallel in $L$. Assume that we have two pairs of parallel joints, we can freeze another $P$-joint of $L$ different from $j_n$ and, in a similar way, we conclude that the directions of all three $P$-joints are same and two of them are neighbouring joints by \cref{p6result}. Then $L$ must be degenerate which is a contradiction. Therefore, in this case, all axes of $R$-joints are parallel in $L$ and there are at least two $P$-joints such that their directions are not parallel to the plane which is perpendicular to the axes of $R$-joints.

\textbf{Case IV: $L$ has $k=4$ or more $P$-joints}. Since the spherical projection $L_s$ has $7-k\le 3$ revolute joints, it is degenerate: all axes are coinciding. Therefore, all rotation axes are parallel and at least two $P$-joints are not perpendicular to the $R$-joints.

%Then all rotation axes are parallel. Because a spherical linkage of $L$ with $6-r\le 3$ revolute joints is necessarily degenerate: if all three (or fewer) joints are actually moving, then all axes are coinciding. Therefore, in this case, all axes of $R$-joints are parallel in $L$ and there are at least two $P$-joints such that their directions must be not parallel to the plane which is perpendicular to the axes of $R$-joints.
\end{proof}

\begin{Thm}\label{pnresult}
% Any $6-$linkage of mobility $2$ or higher containing at least one prismatic joint has parallel revolute axes.
Let $L$ be an $n$-linkage, where $n>6$, of mobility $n-4$  with at least one  $P$-joint and other $R$-joints. Then all axes of $R$-joints are parallel and the number of $P$-joints is at least 2.
\end{Thm}
\begin{proof}
We prove this result by induction on the number of active joints of $L$. The base of the induction is \cref{p7result}.

Fix $n \geq 8$. Suppose that all axes of the $R$-joints in any $(n-k)$-linkage of mobility $n-5$ with at least one $P$-joint  are parallel, where $1\leq k\leq 2$ by \cref{n4fr}. The number of $P$-joints is at least two when $k=1$, and at least two $P$-joints are not perpendicular to the $R$-joints. All $P$-joints are perpendicular to the $R$-joints when $k=2$. Let us freeze an $R$-joint (say $j_{n}$) of $L$ such that one $P$-joint is always active. We get an $(n-l)$-linkage $L'$ with mobility at least $n-5$, where  $1\leq l\leq 2$ by \cref{n4fr}.  %Suppose that all axes of any $(n-k)$-linkage of mobility $n-5$ with at least one $P$-joint  are parallel, where $1\leq k\leq 2$ by \cref{n4fr}, the number of $P$-joints is at least 2 when $k=1$ and the directions of $P$-joints are perpendicular to the direction of the axes  of $R$-joints when $k=2$. Then, freezing an $R$-joint (say $j_{n}$) of $L$ such that one $P$-joint is always active, we get an $(n-l)$-linkage $L'$ with mobility at least $n-5$, where  $1\leq l\leq 2$. See \cref{n4fr}.  

 \textbf{Case I: $l=1$.} By strong induction hypothesis, all axes of $R$-joints of $L'$  (hence of $L$) are parallel and the number of $P$-joints is at least 2 in $L'$ (hence in $L$).

  \textbf{Case II: $l=2$.} Let $j_m$ be the other frozen joint. By strong induction hypothesis, all axes of $R$-joints of $L'$  are parallel and the directions of $P$-joints are perpendicular to the direction of the axes  of $R$-joints. 
	\begin{itemize}
		\item \textbf{If the joint $j_m$ is a $P$-joint}, then the axis of the frozen $R$-joint $j_n$ must be parallel to all axes of $R$-joints of $L'$. Hence, all axes of $R$-joints of $L$  are parallel. As $P$-joints are perpendicular to the $R$-joints in $L'$ (hence of R-joints in $L$), the  $P$-joint $j_m$ must be perpendicular to the $R$-joints in $L$. Then the mobility of $L$ is $n-3$, which is a contradiction. 
\item \textbf{If the joint $j_m$ is an $R$-joint}, then the axes of the frozen $R$-joints $j_m, j_n$ must be parallel. We can freeze a $P$-joint of $L$ such that $j_m$ and $j_n$ are not frozen simultaneously, and the resulting linkage we call $L''$. If $L''$ has a $P$-joint, in a similar way, we conclude that all axes of $R$-joints in $L''$ are parallel. Then all axes of $R$-joints in $L$ are parallel. Then the mobility of $L$ is $n-3$, which is a contradiction. Hence $L''$ has no $P$-joint, by \cref{rnresult}, we conclude that all axes of $R$-joints in $L''$ are parallel. Then all axes of $R$-joints in $L$ are parallel. Then the mobility of $L$ is $n-3$, which is a contradiction.
	\end{itemize} \end{proof}

\begin{Thm}\label{hnresult}
% Any $6-$linkage of mobility $2$ or higher containing at least one prismatic joint has parallel revolute axes.
Let $L$ be an $n$-linkage, $n\geq 7$, of mobility $n-4$ and with at least one  $H$-joint. Then all axes of $R$- and $H$-joints are parallel.
\end{Thm}
\begin{proof}
By \cref{rem:Lr}, we know that the mobility of $L_r$ is greater than or equal to the mobility of $L$. Then the mobility of $L_r$ is at least $n-4$. Furthermore, if the rotation joints of $L_r$ are concurrent, then they must be parallel because at least one $R$-joint comes from an $H$-joint. Therefore, by \cref{rnresult} and \cref{pnresult}, all axes of $L_s$ are equal, hence all axes of $H$- and $R$-joints of $L$ are parallel;
this is possible, and the motion of each joint belongs to the same Sch{\"o}nflies motion subgroup of $SE(3)$.

% We prove this result by induction on the number of active joints.
% 
% For $n = 7$, it is true by \cref{h7result}. 
% 
%  Fix $n \geq 8$. Suppose that all axes of $R$- and $H$-joints  are parallel for any $(n-k)$-linkage with at least one  $H$-joint, and $1\leq k\leq 2$, and the mobility is $n-5$. Let $L$ be an $n$-linkage of mobility $n-4$ with at least one  $H$-joint. Then, freezing an $R$-joint (or $H$-joint if there is no $R$-joint) $j_{s}$ of $L$ such that a helical joint is not frozen simultaneously. We get an $(n-k)$-linkage $L'$ with mobility $n-5$. By strong induction hypothesis, all axes  of $R$- and $H$-joints of $L'$ are parallel. Assume that  $k=1$. The axis of the frozen joint $j_s$ must be parallel to other axes by spherical projection. Hence, all axes of $R$- and $H$-joints are parallel in $L$. Assume that  $k=2$. There is another joint being frozen say $j_r$.  If the joint $j_r$ is a  $P$-joint, then we conclude that the axis of $j_{s}$ must be parallel to the axes of $R$- and $H$-joints in $L'$. Then all axes of $R$- and $H$-joints are parallel. If  the joint $j_r$ is not a  $P$-joint, then the two axes of $j_s$ and $j_r$ must be parallel to each other. Assume that there are two groups of parallel axes. By comparing with \cref{pnresult} and \cref{rnresult}, this is impossible when the mobility of $L_r$ is at least $n-4$. Therefore, all axes of $R$- and $H$-joints are parallel in $L$. 
\end{proof}

% \textcolor[rgb]{1,0,0}{Add thm and proofs}

\section{\texorpdfstring{$nR$}{TEXT}-linkages with mobility \texorpdfstring{$n-5$}{TEXT}}\label{sec:n-5}

In this section we present necessary conditions that the geometry of a paradoxical $nR$-linkage with mobility $n-5$ must satisfy, where $n>6$.

\begin{Lem} \label{lem:7r}
Let $L$ be a $7R$-linkage of mobility 2 with no parallel axes and no Bennett conditions are satisfied. Let $\beta$ be a bond. Then, up to symmetry, we have $A_{\beta}=(j_1,j_4)$. %$A_{\beta}=\{j_1,j_2,j_5\}$  
\end{Lem}

\begin{proof}
We exclude all the possibilities where $|A_{\beta}| \geq 3$ up to symmetry. Notice that whenever $A_{\beta}$ is a chain there is at least one pair of parallel axes, see \cref{lem:chain}. 
\begin{enumerate}
	\item $|A_{\beta}| \geq 6$: This is clearly not possible since in this case $A_{\beta}$ is a chain. 
	\item $|A_{\beta}| = 5$: There are two cases up to symmetry: $A_{\beta}=(j_1,j_2,j_3,j_4,j_6)$ and $A_{\beta}=(j_1,j_2,j_3,j_5,j_6)$. Suppose that $A_{\beta}=(j_1,j_2,j_3,j_4,j_6)$: By assumption there are no parallel axes. By \cref{lem:exabc}, it follows that $(i-h_4)(t_5-h_5)(i-h_6)=0$, 
	%and we have a Bennett condition between 
	and $(h_4,h_5,h_6)$ is a Bennett triple.
	%\begin{align*}
	%(i-h_4)(t_5-h_5)(i-h_6)&=0 \quad \text{or} \\
	%	(i-h_6)(t_7-h_7)(i-h_1)&=0 \quad \text{or} \\
	%		(i-h_4)(t_5-h_5)(i-h_6)&\in E \quad \text{and} \quad  (i-h_6)(t_7-h_7)(i-h_1)\in E.
	%\end{align*}
%	The first two equations are not satisfied since otherwise, by \cref{lem:2sols}, the joint $h_6$ would be paralell to $h_4$ or to $h_1$, respectively. For the same reasons, if the third equation is satisfied, $h_6$ would be parallel to both $h_4$ and $h_6$. 
The argument is the same when $A_{\beta}=(j_1,j_2,j_3,j_5,j_6)$ and we omit it.
	
	\item $|A_{\beta}|=4$: There three cases up to symmetry: $A_{\beta}=(j_1,j_2,j_3,j_5)$, $A_{\beta}=(j_7,j_1,j_4,j_5)$ and $A_{\beta}=(j_7,j_1,j_3,j_5)$. In the first two cases, \cref{lem:exabc} implies that %there is a Bennett condition in 
	the linkage has a Bennett triple. Suppose $A_{\beta}=(j_7,j_1,j_3,j_5)$. If $h_1$ and $h_7$ are not parallel, then 
	\[
	(i-h_1)(t_2-h_2)(i-h_3)(t_4-h_4)(i-h_5)=0% \quad \text{and} \quad (i+h_7)(t_6+h_6)(i+h_5)(t_4+h_4)(i+h_3)=0.  
	\]
	Since $L$ has mobility 2, this defines a curve $C \subset \mathbb{P}^1_{t_2} \times \mathbb{P}^1_{t_4}$ by \cref{cor:hart}. By \cref{lem:curve}, %we have a Bennett condition between 
	either $(h_1,h_2,h_3)$ or $(h_3,h_4,h_5)$ is a Bennett triple.

%	which has a non-constant projection to one of the components, say to  $\mathbb{P}^1_{t_2}$. If $(h_1, h_2)$ and $(h_2,h_3)$ are not parallel, it follows by the extended $abc-$lemma that
	%\[
	%(i-h_3)(t_4-h_4)(i-h_5)=0% \quad \text{and} \quad (i+h_7)(t_6+h_6)(i+h_5)(t_4+h_4)(i+h_3)=0.  
	%\]
	%which is equivalent to a Bennett condition on $(h_3,h_4,h_5)$.
	
	\item If $|A_{\beta}|=3$, then the subsequence $A_{\beta}$ does not contain consecutive elements of $L$: Indeed, suppose $A_{\beta}= (j_1,j_2,j_5 )$. Since $h_1$ and $h_2$ are not parallel, we have the conditions 
		\begin{align*}
	C_1 \colon (i-h_2)(t_3-h_3)(t_4-h_4)(i-h_5)&=0\\
	C_2 \colon	(i+h_1)(t_7+h_7)(t_6+h_6)(i+h_5)&=0
	\end{align*}
	At least one of the $C_i$ is a curve since $L$ has mobility 2. Then there is 
	%a Bennett condition 
	a Bennett triple by \cref{lem:curve}.

	Thus $A_{\beta} = (j_1,j_3,j_5)$: By \cref{lem:exabc}, this is only possible if $(i-h_1)(t_2-h_2)(i-h_3) \in E$. If that was the case then, $i-h_5 \in E$ which is impossible.
	
	\item $|A_{\beta}|=2$: The previous arguments apply to exclude bonds such that $A_{\beta}=(j_1,j_2)$. If $A_{\beta}=(j_1,j_3)$ then 
% 	there would be a Bennett condition between 
	$(h_1,h_2,h_3)$ would be a Bennett triple. 
	\end{enumerate}\end{proof}

%\begin{figure}[tbhp]
%\centering
%\tikzstyle{vertex}=[circle,fill=black, minimum size=2pt,inner sep=2pt]
%\begin{tikzpicture}[transform shape]
%\node[vertex](0) at (4.7, 0.7) {};
%\node[vertex](1) at (5.2, -0.5) {};
%\node[vertex](2) at (5.2, -2) {};
%\node[vertex](3) at (4, -3) {};
%\node[vertex](4) at (2.8,-2) {};
%\node[vertex](5) at (2.8,-0.5) {};
%\node[vertex](6) at (3.3, 0.7) {};
%\begin{scope}[every path/.style={-}, every node/.style={inner sep=0.5pt}]
 %      \draw (6) edge[very thick]  	node[above] {$j_7$} (0);
  %     \draw (0) edge[draw=red, very thick]  	node[right] {$\ j_1$} (1)  ;
	%		 \draw (1) edge[very thick]  	node[right]{$\ j_2$} (2);
	%		 \draw (2) edge[very thick]						node[right, below]{$\quad j_3$}(3) ;
  %     \draw (3) edge[draw=red, very thick]  	node[left, below] {$j_4\quad $}(4); 
	%		 \draw (4) edge[very thick]						node[left] {$j_5\ $}(5);
	%		 \draw (5) edge[very thick]						node[left] {$j_6\ $}(6);
		%	 \draw (0) edge[dashed, bend left] 		node {} (4);
%\end{scope}
%\end{tikzpicture}
%\caption{The only possible bond diagram (up to symmetry) for a 7R with mobility at least 2 and no concurrent axes or Bennett conditions.} 
%\label{fig:7Rm2}
%\end{figure} 

\begin{Thm} \label{lem:4cond}
Let $L$ be a mobile $7R$-linkage with mobility at most 2 and with no concurrent axes or Bennett triples. Then, $L$ has mobility 1. 

%Then there must be $3$ consecutive joints (w.l.o.g., we assume that they are joints $1,2,3$) such that $L_{123}=4$ or $L_{123}=6$.
\end{Thm}  

\begin{proof}
By \cref{lem:7r}, joints $j_1$ and $j_2$ are not simultaneously in $A_{\beta}$. Hence, the projection
 \begin{align*}
  \pi\colon \overline{K_L} &\rightarrow \mathbb{P}^1\times \mathbb{P}^1 \\ 
          (t_1,\ldots,t_7) &\mapsto (t_1,t_2) 
   \end{align*}
is not surjective and the image is contained in a curve. It follows that $t_1$ and $t_2$ are algebraically dependent. In the same way $t_2$ and $t_3$ are algebraically dependent as well and so on and, therefore, all $t_i$ are so. It follows that the mobility of $L$ is 1.
\end{proof}

We now explain how to generalise this result to any number of joints.

\begin{Thm} \label{thm:nrpartial}
Let $L$ be an $nR$-linkage with $n\geq 7$ and no concurrent axes or Bennett triples. Then $L$ has mobility at most $n-6$ (i.e., it does not have paradoxical mobility).
\end{Thm}
\begin{proof}
We prove this result by induction on the number of active joints where the base case is \cref{lem:4cond}. Fix $n \geq 8$. We freeze a joint in $L$ and we have an $(n-1)R$-linkage, say $L'$. We proceed by looking at the number of joints that get frozen as a result (in each case we are going to prove that $L'$ has mobility at most $n-7$).

\textbf{Case I: One joint gets frozen.} We claim that $L'$ has no concurrent axes or Bennett triples. Otherwise, by \cref{lem:fr}, we get the contradiction that $L$ has concurrent axes or Bennett triples. By the induction hypothesis, $L'$ has mobility at most $n-7$. 

\textbf{Case IIa: Two joints get frozen and $n=8$.} The linkage $L'$  is a $6R$-linkage. By \cref{classificationthm:6rm2}, the mobility of $L'$ is at most 1. Otherwise, there would be at least one Bennett triple in $L$. 

\textbf{Case IIb: Two joints get frozen and $n\ge 9$.} The linkage $L'$ cannot have all $n-2$ axes concurrent. By \cref{rnresult}, it follows that $L'$ has mobility at most $n-7$.

\textbf{Case IIIa: Three joints get frozen and $n=8$.} The linkage $L'$ cannot have all $5$ axes concurrent. By \cref{prop:SE2}, it follows that $L'$ has mobility at most 1.

\textbf{Case IIIb: Three joints get frozen and $n=9$} (this case is very similar to Case IIIa). The linkage $L'$ cannot have all $6$ axes concurrent. By \cref{prop:SE2}, it follows that $L'$ has mobility at most 2.

\textbf{Case IIIc: Three joints get frozen and $n\ge 10$.} The linkage $L'$ cannot have all $n-3$ axes concurrent. By \cref{rnresult}, it follows that $L'$ has mobility at most $n-7$.
 
\textbf{Case IV: Four or more joints get frozen.} Then the linkage $L'$ is an $(n-s)R$-linkage. By \cref{prop:maxmob}, it follows that the mobility of $L'$ is at most $n-s-3\le n-7$.

In all cases, we have shown that the mobility of $L'$ is at most $n-7$. This implies that $L$ has mobility at most $n-6$.
\end{proof}

\section*{Acknowledgements} The first author would also like to express his gratitute to Hamid Abban for his encouragement
and for putting him in contact with the research group led by Josef Schicho. He was supported by EPSRC grant ref. EP/V048619/1 and by the Austrian Science Fund (FWF): W1214-N15, project DK9 while staying at RICAM. This research was partially supported by the Austrian Science Fund (FWF): W1214-N15, project DK9. The research was funded by the Austrian Science Fund (FWF): P 31061

\end{document}